\documentclass[nojss]{jss}
\usepackage{orcidlink,thumbpdf,lmodern}

\usepackage{framed}

\usepackage{graphicx} % Required for inserting images
\usepackage{xcolor} %colored characters
\usepackage{CJKutf8} 
\usepackage{graphicx} 
\usepackage{float}
\usepackage{amssymb} 
\usepackage{amsmath} 
\usepackage{geometry} 
\usepackage{makecell} 
\usepackage{hyperref} 
\usepackage{url} 
\usepackage{subcaption}
\usepackage{listings} 
\usepackage{algorithm} 
\usepackage{algpseudocode} 
\usepackage{verbatim} 
\usepackage{multirow} 
\usepackage{enumitem} 
\newcommand{\class}[1]{`\code{#1}'}
\newcommand{\fct}[1]{\code{#1()}}

\author{
        Heyang Ji~\orcidlink{0009-0001-7494-7227}\\Indiana University
   \And Ufuk Beyaztas\\Marmara University
   \And Nicolas Escobar-Velasquez\\Indiana University
   \AND Yuanyuan Luan\\Indiana University
   \And Xiwei Chen\\Indiana University
   \AND Mengli Zhang\\Shanghai University of Finance and Economics
   \And Roger Zoh\\Indiana University
   \AND Lan Xue\\Oregon State University
   \And Carmen Tekwe\\Indiana University
   }
\Plainauthor{Heyang Ji, Ufuk Beyaztas, Nicolas Escobar-Velasquez, Yuanyuan Luan, Xiwei Chen, Mengli Zhang, Roger Zoh, Lan Xue, Carmen Tekwe}

\title{\pkg{MECfda}: An \proglang{R} Package for Bias Correction Due to Measurement Error in Functional and Scalar Covariates in Scalar-on-Function Regression Models
}
\Plaintitle{MECfda: An R Package for Bias Correction Due to Measurement Error in Functional and Scalar Covariates in Scalar-on-Function Regression Models
}
\Shorttitle{Introduction to \proglang{R} Package \pkg{MECfda}}

\Abstract{
Functional data analysis (FDA) deals with high-resolution data recorded over a continuum, such as time, space or frequency. Device-based assessments of physical activity or sleep are objective yet still prone to measurement error. We present \pkg{MECfda}, an \proglang{R} package that (i) fits scalar-on-function, generalized scalar-on-function, and functional quantile regression models, and (ii) provides bias-corrected estimation when functional covariates are measured with error. By unifying these tools under a consistent syntax, \pkg{MECfda} enables robust inference for FDA applications that involve noisy functional data.
}

\Keywords{\pkg{MECfda}, functional data, measurement error, scalar-on-function regression, \proglang{R}}
\Plainkeywords{MECfda, functional data, measurement error, scalar-on-function regression, R}

\Address{
  Heyang Ji\\
  Department of Epidemiology and Biostatistics\\
  Ph.D. candidate\\
  Indiana University\\
  1025 E. Seventh Street, Suite 111\\
  Bloomington, Indiana, United States, 47405-7109\\
  E-mail: \email{hj41@iu.edu}\\
}

\begin{document}

\section[Functional data and scalar-on-function regression]{Functional data and scalar-on-function regression} \label{sec:FDA}

\subsection{Functional data}

Functional data analysis is commonly used 
to analyze high-dimensional data that appear as functions or images 
\citep{wang2016functional,ramsay2002applied,ramsay1991some}. 
Functional data analysis can be used to analyzed data collected 
continuously or frequently over a given time period that appear as 
complex high dimensional functions of time. 
When assessing how both functional and scalar-valued covariates 
influence scalar-valued outcomes, 
scalar-on-function regression models may be used. 
Self-reported measures, such as dietary intake,
are well known to be prone to measurement error \citep{carroll2006measurement},
and recent studies have indicated that even the relatively more objective data 
obtained from wearable devices is prone to measurement error
\citep{crainiceanu2009generalized,tekwe2019instrumental}. 
It has been demonstrated that similar to scalar-valued covariates prone 
to measurement error, the failure to correct for biases due to measurement error 
associated with functional data also leads to biased estimates.

We develop an 
\proglang{R} \citep{R-base}
package \pkg{MECfda}
for solving scalar-on-function regression models including
multi-level generalized scalar-on-function regression models 
and functional quantile linear regression models
measurement error corrections using various bias reduction techniques 
in these models.

\paragraph{Data structure.}
Let $x_{ij}=f_i(t_j)$ denote the value of function $f_i$ measured at time
$t_j$.  The $n\times m$ matrix $(x_{ij})$ contains one function per row and
one time point per column.

\subsection{Scalar-on-function linear regression models}

The general SoFR model is
\begin{equation}
  T\!\bigl(F_{Y_i\mid X_i,Z_i}\bigr)=
  \sum_{l=1}^L \int_{\Omega_l} \beta_l(t) X_{li}(t)\,dt + (1,Z_i^\top)\gamma,
  \label{eq:sofr-general}
\end{equation}
where $Y_i$ is a scalar response,
$X_{li}(\cdot)\in L^2(\Omega_l)$ are functional covariates with coefficients
$\beta_l(\cdot)$, $Z_i$ are scalar covariates, and $T(\cdot)$ is a statistical functional.

\paragraph{Terminology.}
In the context of \emph{functional data analysis} (FDA), the word \emph{functional} is used
\emph{as an adjective}: it qualifies the noun "data" to indicate that each
observation \(X_i\) is a real-valued function \(t \mapsto X_i(t)\) defined on
a compact domain~\(\Omega\).  
By contrast, in the context of \emph{statistical functionals} 
the same word \emph{functional} is a \emph{noun}.
And a statistical functional is defined as a mapping
\(T\colon\mathcal{F}\to\mathbb{R}\), where
\(\mathcal{F}\) is a class of probability distributions.  \citep{wasserman2006all}
Statistical functionals represent quantifiable properties of probability distributions, e.g., mean or quantile)

We focus on two special cases supported in \pkg{MECfda}:
\begin{enumerate}
    \item \textbf{Generalized SoFR:}  
    This model is formulated as
    \begin{equation}
        \label{eq: GLM SoFR}
        T\bigl(F_{Y_i\mid X_i,Z_i}\bigr) = g\Bigl\{\mathbb{E}(Y_i\mid X_i,Z_i)\Bigr\} = g\left\{\int_\mathbb{R} y dF_{Y_i|X_i,Z_i}(y)\right\},
    \end{equation}
    where $g(\cdot)$ is a strictly monotonic link function analogous to that used in generalized linear models.

    \item \textbf{Quantile SoFR:}  
    This model is expressed as
    \begin{equation}
        \label{eq: QR SoFR}
        T\bigl(F_{Y_i\mid X_i,Z_i}\bigr) = Q_{Y_i\mid X_i,Z_i}(\tau) = F_{Y_i\mid X_i,Z_i}^{-1}(\tau),
    \end{equation}
    where $\tau \in (0,1)$ denotes the quantile of interest.
\end{enumerate}

\subsection{Basis expansion} 

Let $\{\rho_k\}_{k=1}^{\infty}$ be a complete basis of $L^2(\Omega)$.
For $\beta\in L^2(\Omega)$,
$\beta(t)=\sum_{k=1}^{\infty} c_k \rho_k(t)$.
Truncating at $p$ terms gives
\[
  \int_\Omega\beta(t)X_i(t)\,dt
  \approx \sum_{k=1}^{p} c_k
           \underbrace{\int_\Omega X_i(t)\rho_k(t)\,dt}_{b_{ik}},
\]
so that each $b_{ik}$ becomes a scalar covariate and only
$c_1,\dots,c_p$ need to be estimated.
The number of basis functions $p=p(n)$ typically grows slowly with $n$.

In practice, we may not necessarily use the truncated complete basis of the $L^2$ function space. 
Instead, we can use a finite sequence of linearly independent functions as the basis functions. 
Common choices include the Fourier, B-spline, and eigenfunction
(functional principal component, FPC) bases; all are supported in
\pkg{MECfda}.
(see Appendix~\ref{def basis function} for the full recurrence of the definition of these basis). 

When analytical forms are unavailable, \pkg{MECfda}
stores the numerical evaluations $\rho_k(t_j)$ in an $m\times p$
matrix for efficient computation.
 
\section{Measurement error problems in scalar-on-function regression}\label{sec:ME}

Data collected in real-world settings often include measurement error, 
especially functional data. 
Measurement error in a data set may result in estimation bias. 
The \pkg{MECfda} package also provides bias correction estimation method functions for
certain linear regression models for use with data containing measurement error.

Measurement error is an important problem in functional data analysis. 
Functional data are often contaminated by measurement error. 
For example, the physical activity data collected by wearable devices. 
Measurement error induces biased parameter estimates. 
There have been some methods proposed to correct the bias due to measurement error. 
The package \pkg{MECfda} provide some functions to realize some of the method to 
address the measurement error in scalar-on-function regression.

\subsection{Mixed effect model (MEM) based method by Luan et al.}\label{subsec:ME MEM}

Wearable monitoring devices permit the continuous monitoring of biological processes, 
such as blood glucose metabolism, and behaviors, such as sleep quality and physical activity.  
Continuous monitoring often collects data in 60-second epochs over multiple days,
resulting in high-dimensional multi-level longitudinal curves that are best described and analyzed as multi-level functional data.  
Although researchers have previously addressed measurement error in scalar covariates prone to error,
less work has been done for correcting measurement error 
in multi-level high-dimensional curves prone to heteroscedastic measurement error. 
Therefore, \citet{luan2023scalable} proposed mixed-effects model-based (MEM) methods 
for bias correction due to measurement error in multi-level functional data from 
the exponential family of distributions that are
prone to complex heteroscedastic measurement error. % \citep{luan2023scalable}

The working model can be written as
\begin{align*}
  &g(E(Y_i\mid X_i,Z_i)) = \int_{\Omega} \beta(t) X_{i}(t) dt + (1,Z_i^\top)\gamma,\\
  &h(E(W_{ij}(t)\mid X_i(t))) = X_i(t),\\
  &X_i(t) = \mu_x(t) + \varepsilon_{xi}(t),
\end{align*}
where $Y_i$ is a scalar response, $Z_i$ is a vector of error-free scalar covariates, $X_i(t)$ is the latent functional predictor, and $W_{ij}(t)$ is its $j$-th noisy replication for subject $i$.

\paragraph{Assumptions}
\begin{enumerate}
    \item $Y_i\mid X_i,Z_i\sim EF(\cdot)$, where $EF$ refers to an exponential family distribution. 
    \item $g(\cdot)$ and $h(\cdot)$ are known to be strictly monotone, twice continuously differentiable functions.
    \item $\operatorname{Cov}\{X_i(t),W_{ij}(t)\} \neq 0$. 
    \item $W_{ij}(t)\mid X_i(t)\sim EF(\cdot)$.
    \item $f_{Y_i\mid W_{ij}(t),X_i(t)}(\cdot) = f_{Y_i\mid X_i(t)}(\cdot)$, where $f$ refers to a probability density function.  
    \item $X_i(t)\sim GP\{\mu_x(t),\Sigma_{xx}\}$, where $\mathrm{GP}$ refers to the Gaussian process. 
\end{enumerate}

These assumptions allow for 
i) a more general specification of the distributions of error-prone functional covariates compared to current approaches that often entail normality assumptions for these observed measures and
ii) a nonlinear association between the true measurement and the observed measurement prone to measurement error.

Estimation proceeds in two stages.  
Stage 1 fits employ point-wise (\code{UP\_MEM}) or multi-point-wise (\code{MP\_MEM}) procedure for fitting the multi-level functional MEM approach, 
avoiding the need to compute complex and intractable integrals that would be required in traditional approaches for reducing biases due to measurement error in multi-level functional data. 
They treat the fixed intercept as the mean trajectory $\mu_x(t)$ and the subject-specific random intercept $\varepsilon_{xi}(t)$ as deviations, thereby yielding a bias-corrected proxy $\widehat X_i(t)$.  
Stage 2 substitutes $\widehat X_i(t)$ for $X_i(t)$ in the scalar-on-function regression and fits the model.  
This two-stage MEM strategy achieves bias reduction while avoiding the high-dimensional integrals that plague traditional likelihood-based corrections for multilevel functional measurement error.
Functions \fct{ME.fcRegression\_MEM} in \pkg{MECfda} implement the full workflow.

\subsection{Simulation extrapolation (SIMEX) correction by Tekwe et. al.}\label{subsec:ME SIMEX}

Wearable sensors have become the primary source of objective, high-frequency
activity curves.  Although more reliable than self-reports, the recorded
signals \(W_i(t)\) are still contaminated by calibration drift, device
placement, and proprietary processing.  When scalar outcomes—such as body-mass
index (BMI)—are related to these error-prone curves, standard regression
delivers biased inference.

Quantile regression is a tool developed to model complex relationships between a set of covariates and quantiles of an outcome. 
Compared to traditional linear regression approaches, 
quantile regression does not assume a specific residual distribution and is robust to outliers.

\citet{tekwe2022estimation} developed a two-stage
instrumental-variable Simulation–Extrapolation (SIMEX)
procedure for scalar-on-function quantile regression.

The working model can be written as
\begin{align*}
  &Q_{Y_i\mid X_i,Z_i}(\tau) = \int_{\Omega} \beta(\tau,t) X_{i}(t) dt +  (1,Z_i^\top)\gamma(\tau),\\
  &W_i(t) = X_i(t) + U_i(t),\\
  &M_i(t) = \delta(t) X_i(t) + \eta_i(t),
\end{align*}

where \(Y_i\) and the scalar covariates \(Z_i\) are measured without error,
\(X_i(t)\) is the latent true curve, \(W_i(t)\) its noisy surrogate, and
$M_i(t)$ is a measured functional instrumental variable.

\paragraph{Assumptions}
\begin{enumerate}
  \item \(\operatorname{Cov}\!\{X_i(t),U_i(s)\}=0\),
  \item \(\operatorname{Cov}\!\{M_i(t),U_i(s)\}=0\),
  \item \(\mathbb{E}\!\bigl\{W_i(t)\mid X_i(t)\bigr\}=X_i(t)\),
  \item \(U_i(t)\sim\mathrm{GP}\{0,\Sigma_{uu}\}\),\quad
        \(\forall\,t,s\in[0,1]\), where $\mathrm{GP}$ refers to the Gaussian process.
\end{enumerate}

\paragraph{Two-stage algorithm}
\begin{enumerate}
  \item \textbf{IV calibration}\;—
        use \(M_i(t)\) to estimate the error covariance
        \(\Sigma_{uu}\).
  \item \textbf{SIMEX correction}\;—
        add pseudo-noise to \(W_i(t)\), refit the quantile model
        over a grid of noise levels, and extrapolate back to the
        error-free scenario to obtain bias-corrected
        \(\hat\beta(\tau,t)\) and \(\hat\gamma(\tau)\).
\end{enumerate}

This IV-SIMEX scheme inherits the robustness of quantile regression,
does not rely on Gaussian residuals, and effectively removes the
attenuation bias induced by functional measurement error.
Functions \fct{ME.fcQR\_IV.SIMEX} in \pkg{MECfda} implement the full workflow.

\subsection{Corrected loss score method by Zhang et al.}\label{subsec:ME CLS}

\citet{zhang2023partially} proposed a corrected-loss approach for partially functional linear quantile regression models when the functional covariate is contaminated with measurement error.
This method is designed to analyze scalar responses (e.g., BMI) influenced by both scalar covariates and functional predictors, the latter of which are only indirectly observed with additive measurement errors---a common situation when using wearable devices to monitor physical activity.

The working model is specified as
\begin{align*}
  &Q_{Y_i\mid X_i,Z_i}(\tau) = \int_0^1 \beta(\tau,t) X_i(t)\,dt + Z_i^\top\theta(\tau), \\
  &W_i(t) = X_i(t) + U_i(t),
\end{align*}
where \(Y_i\in\mathbb{R}\) is the scalar response, \(Z_i\in\mathbb{R}^p\) is a
vector of error-free covariates, \(X_i(t)\in L^2[0,1]\) is the latent
functional covariate, and \(W_i(t)\) is its observed error-prone version.
The measurement error process \(U_i(t)\) is assumed to be centered and
independent of both the response and covariates.

\paragraph{Assumptions}
\begin{enumerate}
  \item \(\{(Y_i, X_i(t), Z_i, U_i(t))\}_{i=1}^n\) are i.i.d.\ samples,
  \item \(U_i(t)\sim \mathrm{GP}(0,\Sigma_{uu})\) and independent of \((X_i(t), Z_i, Y_i)\),
  \item \(X_i(t)\) admits a basis expansion with eigenvalues \(\kappa_j \asymp j^{-\alpha_x}\) for some \(\alpha_x > 1\),
  \item \(\beta(\tau,t)\) admits a basis expansion with coefficients satisfying \(|b_j(\tau)| \leq C j^{-\beta}\), where \(\beta > \alpha_x/2 + 1\),
  \item \(X_i(t)\) and \(U_i(t)\) are both mean-square continuous in \(t\),
  \item The design matrix formed by principal components of \(X_i(t)\) and \(Z_i\) has bounded eigenvalues,
  \item The conditional density of \(Y_i\) given \(X_i(t), Z_i\) is continuous and strictly positive near the conditional quantile.
\end{enumerate}

\paragraph{Corrected loss estimation}
\begin{enumerate}
  \item \textbf{Basis expansion}\;—
        apply functional principal component analysis (FPCA) to represent
        \(X_i(t)\), \(\beta(t)\), and \(U_i(t)\) via orthonormal eigenbasis.
  \item \textbf{Loss correction}\;—
        define a smooth surrogate \(\rho_h(\cdot)\) of the quantile check loss
        and construct a corrected loss function \(\rho_h^*(\cdot, \sigma^2)\) to
        account for the variance inflation due to measurement error.
  \item \textbf{Optimization}\;—
        minimize the corrected loss
        \[
        \sum_{i=1}^n \rho_h^*\left(Y_i - W_i^\top b - Z_i^\top\theta,\, b^\top \Sigma_u b\right)
        \]
        with respect to the coefficients \(b\) and \(\theta\), where \(\Sigma_u\) is estimated from replicate observations of \(W_i(t)\).
\end{enumerate}

This corrected-loss approach achieves consistent estimation without requiring knowledge
of the response distribution and is robust to correlation in the measurement error process.
Functions \fct{ME.fcQR\_CLS} in \pkg{MECfda} implement the full estimation workflow.

\subsection{Instrumental variable based method by Tekwe et al. }\label{subsec:ME IV}

\citet{tekwe2019instrumental} developed an instrumental–variable
Generalized Method of Moments (IV–GMM) estimator for scalar–on–function
regression with functional measurement error.

The working model is
\begin{align*}
  &Y_i \;=\; \int_{\Omega} \beta(t)\,X_i(t)\,dt \;+\;\varepsilon_i,\\
  &W_i(t) \;=\; X_i(t) + U_i(t),\\
  &M_i(t) \;=\; \delta(t)\,X_i(t) + \omega_i(t),
\end{align*}
where \(Y_i\) is measured without error, \(X_i(t)\) is the latent true
curve, \(W_i(t)\) its noisy surrogate, and \(M_i(t)\) a measured
functional instrumental variable.

\paragraph{Assumptions}
\begin{enumerate}
  \item \(\operatorname{Cov}\!\{X_i(t),U_i(s)\}=0\) for all
        \(t,s\in\Omega\) (classical measurement error);
  \item \(\operatorname{Cov}\!\{M_i(t),U_i(s)\}=0\) and
        \(\operatorname{Cov}(M_i(t),\varepsilon_i)=0\)
        (instrument exogeneity);
  \item \(\operatorname{Cov}\!\{M_i(t),X_i(s)\}\) is full rank
        (identifiability);
  \item \(\beta(t)\) is \((p{+}1)\)-times continuously differentiable on
        \(\Omega\);
  \item The processes \(X_i,U_i,\omega_i\) have bounded fourth moments,
        and \(\varepsilon_i\) has finite variance.
\end{enumerate}

\paragraph{Estimation algorithm}
\begin{enumerate}
  \item \textbf{Basis projection}—Approximate
        \(\beta(t)\approx\sum_{k=1}^{K_n}\gamma_k b_k(t)\) and project
        \(X_i,W_i,M_i\) onto the same B-spline basis to obtain the
        score vectors \(\mathbf X_i,\mathbf W_i,\mathbf M_i\in\mathbb{R}^{K_n}\).
  \item \textbf{GMM solution}—Form the empirical moment matrices
        \(\widehat{\boldsymbol\Omega}_{WM}=n^{-1}\sum_i
         \mathbf W_i\mathbf M_i^{\top}\) and
        \(\widehat{\boldsymbol\Omega}_{MY}=n^{-1}\sum_i
         Y_i\mathbf M_i\); solve
        \[
          \hat{\boldsymbol\gamma}
          =
          \bigl(\widehat{\boldsymbol\Omega}_{WM}^{\top}
                 \widehat{\boldsymbol\Omega}_{WM}\bigr)^{-1}
          \widehat{\boldsymbol\Omega}_{WM}^{\top}
          \widehat{\boldsymbol\Omega}_{MY}.
        \]
  \item \textbf{Curve reconstruction}—Recover the slope function by
        \(\hat\beta(t)=\sum_{k=1}^{K_n}\hat\gamma_k\,b_k(t)\).
\end{enumerate}

This IV–GMM scheme avoids explicit estimation of the measurement error
covariance, accommodates non–Gaussian or heteroscedastic errors, and
achieves the minimax‐optimal \(L^2\) convergence rate under the above
regularity conditions.
Function \fct{ME.fcLR\_IV} in the \pkg{MECfda} package
implements the complete workflow.

\section{Software for solving the models}\label{sec:software}

\paragraph{From methodology to implementation.} Section~\ref{sec:FDA}  and ~\ref{sec:ME} described the measurement-error-aware scalar-on-function regression framework that motivates \pkg{MECfda}. 
The present section explains how these statistical ideas are translated into a coherent \proglang{R} interface. 
We first outline the overall design philosophy and a typical analysis workflow, before documenting the S4 infrastructure, regression engines, and measurement error modules. 

\subsection{Design philosophy and workflow}

\begin{enumerate}
  \item \textbf{Data abstraction}: Raw matrices of functional observations are wrapped by the \class{functional\_variable} class, ensuring that downstream functions always receive metadata such as domain and grid.
  \item \textbf{Basis expansion layer}: Helper functions of the form \fct{*\_basis\_expansion} convert functional variables to low dimensional coefficient matrices. Any basis object can subsequently be coerced to an \proglang{R} function by \fct{basis2fun} for prediction and plotting.
  \item \textbf{Estimation engines}: High level wrappers (e.g., \fct{fcRegression}, \fct{fcQR}) operate on the coefficient matrices and call established back-end packages such as \pkg{lme4} and \pkg{quantreg}.
  \item \textbf{Measurement error corrections}: Dedicated front ends (prefix \code{ME.}) replace noisy surrogates by unbiased proxies or invoke SIMEX/CLS algorithms before delegating to the core engines.
  \item \textbf{Post-processing}: Methods for parameter extraction, coefficient plotting, prediction, and bootstrap inference are provided for every fitted model.
\end{enumerate}

\begin{table}[tb]
  \centering\small
  \caption{Mapping between statistical methods and \pkg{MECfda} entry points.}\label{tab:map}
  \begin{tabular}{lll}
    \hline
    Method & Equation or section & \proglang{R} function \\
    \hline
    Generalized SoFR & Eq~\eqref{eq: GLM SoFR} & \fct{fcRegression} \\
    Quantile SoFR & Eq~\eqref{eq: QR SoFR} & \fct{fcQR} \\
    MEM substitution & Section~\ref{subsec:ME MEM} & \fct{ME.fcRegression\_MEM} \\
    IV-SIMEX quantile & Section~\ref{subsec:ME SIMEX} & \fct{ME.fcQR\_IV.SIMEX} \\
    CLS quantile & Section~\ref{subsec:ME CLS} & \fct{ME.fcQR\_CLS} \\
    IV two-stage least squares & Section~\ref{subsec:ME IV} & \fct{ME.fcLR\_IV} \\
    \hline
  \end{tabular}
\end{table}

\subsection{S4 classes}

\paragraph{\class{functional\_variable}}\quad
Stores functional observations in a matrix $X=(x_{ij})$ together with the
measurement grid and domain $(t_0,T)$. Methods such as \fct{dim}, \fct{plot},
and the basis-expansion family treat the object as the canonical carrier of
functional covariates.
% \begin{Code}
% fv <- functional_variable(matrix(rnorm(10*24), 10, 24),
%                           domain_start = 0, domain_length = 1,
%                           grid = (0:23) / 24)
% dim(fv)
% \end{Code}

\paragraph{\class{Fourier\_series}}\quad
Encodes $\tfrac{a_0}{2}+\sum_k a_k\cos(2\pi k\cdot)+b_k\sin(2\pi k\cdot)$ on
$[t_0, t_0+T]$ with explicit coefficient slots. It comes with fast evaluation
(\fct{FourierSeries2fun}) and plotting, making it convenient for periodic
effects and smooth approximations.
% \begin{Code}
% fs <- Fourier_series(1, cos = 1, sin = 0, k_cos = 1, k_sin = 1)
% plot(fs)
% \end{Code}

\paragraph{\class{bspline\_basis} / \class{bspline\_series}}\quad
\class{bspline\_basis} defines a knot sequence and polynomial degree for the
B-spline family $\{B_{i,p}\}$, while \class{bspline\_series} stores
coefficients $\{b_i\}$ to form $\sum_i b_i B_{i,p}(x)$. Together they offer
local flexibility and explicit continuity control.
% \begin{Code}
% bb <- bspline_basis(df = 6, degree = 3)
% bs <- bspline_series(runif(6), bb)
% plot(bs)
% \end{Code}

\paragraph{\class{numeric\_basis} / \class{numericBasis\_series}}\quad
Represent any finite set of basis functions numerically on a common grid,
freeing the user from analytic formulas. This is useful for empirical eigen
bases (FPC) or externally supplied wavelets.

\subsection{Core utilities}

\paragraph{\fct{basis2fun}}\quad Generic coercion that turns any series
object into a fast \proglang{R} function $f(t)$, enabling prediction and plotting
with uniform syntax.

\paragraph{\fct{fourier\_basis\_expansion}},
{\fct{bspline\_basis\_expansion}},
{\fct{FPC\_basis\_expansion}},
{\fct{numeric\_basis\_expansion}}\quad Convert a
\class{functional\_variable} into coefficient matrix $C=(c_{ik})$ such that
$X_i(t)\approx\sum_k c_{ik}B_k(t)$; each wrapper returns both $C$ and the
chosen basis object.

\subsection{Regression engines}

\paragraph{\fct{fcRegression}}\quad Fits a (generalized) linear
mixed effect model in Eq~\eqref{eq: GLM SoFR}
via basis expansion, supporting Gaussian, binomial, Poisson, etc. Random
effects in \code{Z} are specified by \pkg{lme4} style formulae.

\paragraph{\fct{fcQR}}\quad Fits a (generalized) linear
mixed effect model in Eq~\eqref{eq: QR SoFR}
via basis expansion. 

\subsection{Measurement error tool set}

\paragraph{\fct{ME.fcRegression\_MEM}}\quad Applies predictor
(\textsc{mem}) substitution $\hat X$ before calling \fct{fcRegression};
supports Gaussian/Poisson error in the surrogate $W$ and optional smoothing of
$\hat X$.
% A overview of the algorithm appears in Section~\ref{subsec:ME MEM}.

\paragraph{\fct{MEM\_X\_hat}}\quad Returns the $\hat X$ (predicted
$X$) used in \fct{ME.fcRegression\_MEM}.

\paragraph{\fct{ME.fcQR\_IV.SIMEX}}\quad Combines instrumental
variables with a SIMEX extrapolation to correct functional quantile regression
when $X$ is noisy and auxiliary $M$ is available.

\paragraph{\fct{ME.fcQR\_CLS}}\quad Implements the corrected-loss
approach for functional quantile regression; automatically selects basis size
$k$ and bandwidth $h$ over user-supplied grids.

\paragraph{\fct{ME.fcLR\_IV}}\quad Two-stage IV estimator for the
linear model $Y_i=\int\beta(t)X_i(t)\,dt+\varepsilon_i$ using $M$ as
instrument, with optional bootstrap confidence intervals.

\subsection*{From functions to full analyses}

The preceding sections have introduced the S4 infrastructure, estimation
engines, and measurement--error utilities exported by \pkg{MECfda}. 
To demonstrate how these programmatic components come together in a real data
analysis, the next section (\S\ref{sec:Example Project}) walks through a complete,
reproducible pipeline on the 2011--2012 NHANES accelerometry cohort.

\section[Example project using MECfda]{Example project using \pkg{MECfda}}\label{sec:Example Project}

The National Health and Nutrition Examination Survey (NHANES) is a program conducted by the National Center for Health Statistics (NCHS) designed to assess the health and nutritional status of adults and children in the United States. It combines interviews and physical examinations to collect comprehensive health-related data from a nationally representative sample. Among its various components, NHANES includes objective physical activity data collected via wearable accelerometers. These devices provide minute-level records of participants’ movement intensity, allowing researchers to evaluate daily activity patterns, sedentary behavior, and compliance with physical activity guidelines. 
In addition to the raw physical activity monitor data provided by NHANES, step‐count summaries derived from the raw signals are available through the
\href{https://physionet.org/content/minute-level-step-count-nhanes/1.0.1/}{\textit{PhysioNet}}
repository ~\citep{koffman2025nhanessteps, koffman2024comparing}.
PhysioNet is an open-access resource funded by the National Institutes of Health (NIH) that provides large-scale physiological and clinical datasets along with open-source tools for biomedical signal analysis \citep{goldberger2000physiobank}. It includes a wide variety of data such as ECG, EEG, blood pressure, and activity data collected from wearable sensors. For NHANES, PhysioNet hosts step count estimates for the 2011-2014 survey cycles, enabling researchers to use high-resolution, minute-level physical activity metrics for epidemiological and public health research.

We analyse the 2011–2012 cohort to investigate how hourly activity intensity predicts body‐mass index (BMI) after adjusting for sex, age, and race.

% An \proglang{R} package \pkg{nhanesA} \citep{nhanesA} has been developed to provide a convenient interface to access publicly available NHANES datasets directly from the CDC website. It allows users to search, download, and import NHANES data files (including demographic, examination, laboratory, etc) into \proglang{R} with simple commands, making it easier for researchers to integrate NHANES data into reproducible \proglang{R} workflows.

\subsection{Data access and pre-processing}\label{subsec:nhanes-prep}

The following pipeline downloads and pre-processes the data.

\paragraph{Step 1: Hour‑level accelerometer records (PAXHR).}  We download the file
\texttt{PAXHR\_G}, which stores hour-level physical activity data of cohort 2011-2012 for up to nine monitoring days.  
In the NHANES Physical Activity Monitor protocol, participants wear the accelerometer continuously for seven full days (midnight to midnight), removing it on the morning of the ninth day.  Hence, hour-level records from day 1 and day 9 may be incomplete and are excluded.  We retain only participants with all 9 days of data and full 24-hour records on days 2–8 \citep{nhanes2011paxhr}.
The result is reshaped into a $n\times24\times7$ array.

\begin{comment}
\begin{CodeChunk}
\begin{CodeInput}
PAXHR = nhanes("PAXHR_G")

PAXHR = PAXHR %>%
  dplyr::group_by(SEQN, PAXDAYH) %>%
  dplyr::arrange(PAXSSNHP, .by_group = TRUE) %>%
  dplyr::mutate(hour_in_day = row_number() - 1) %>%
  dplyr::ungroup()

id_all9days = PAXHR %>%
  dplyr::group_by(SEQN) %>%
  dplyr::summarise(n_days = n_distinct(PAXDAYH),
                   .groups = "keep") %>%
  dplyr::filter(n_days == 9) %>%
  dplyr::pull(SEQN)

PAXHR = PAXHR %>%
  dplyr::filter(SEQN %in% id_all9days)
rm(id_all9days)

id_completeData = PAXHR %>%
  dplyr::filter(as.integer(PAXDAYH) %in% 2:8) %>% 
  dplyr::group_by(SEQN, PAXDAYH) %>% 
  dplyr::summarise(n_hour = n_distinct(PAXSSNHP), 
                   .groups = "drop") %>%  
  dplyr::filter(n_hour == 24) %>%  
  dplyr::group_by(SEQN) %>%
  dplyr::summarise(n_day = n()) %>% 
  dplyr::filter(n_day == 7) %>% 
  dplyr::pull(SEQN)

PAXHR = PAXHR %>%
  filter(SEQN %in% id_completeData, as.integer(PAXDAYH) %in% 2:8)
PAXHR$PAXDAYH = droplevels(PAXHR$PAXDAYH)

PAXHR = PAXHR[,
  colnames(PAXHR) %in% c('SEQN','PAXDAYH','hour_in_day','PAXMTSH')]

PAXHR = PAXHR %>%
  tidyr::pivot_wider(
    names_from = hour_in_day,    
    values_from = PAXMTSH      
  )

PAXHR = PAXHR[order(match(PAXHR$SEQN, id_completeData)), ]
day_list_PA = split(PAXHR[,as.character(0:23)], PAXHR$PAXDAYH)
day_list_PA = lapply(day_list_PA, as.matrix)
PA_array = abind::abind(day_list_PA, along = 3)
\end{CodeInput}
\end{CodeChunk}
\end{comment}

\paragraph{Step 2: Outcome and demographic covariates.}  
The outcome variable, BMI, comes from the body measure data in file \texttt{BMX\_G}.  
The demographic covariates, sex, age, and race are extracted from the demographic data in file \texttt{DEMO\_G}.

\begin{comment}
\begin{CodeChunk}
\begin{CodeInput}
BMX = nhanes("BMX_G")
BMX = BMX[BMX$SEQN %in% id_completeData,
          c('BMXBMI'),drop = FALSE]
BMX = BMX[order(match(BMX$SEQN, id_completeData)), ]
BMX = BMX[,c('BMXBMI'),drop = FALSE]

DEMO = nhanes("DEMO_G")
DEMO = DEMO[DEMO$SEQN %in% id_completeData,
            c('SEQN','RIAGENDR','RIDAGEYR','RIDRETH1')]
DEMO = DEMO[order(match(DEMO$SEQN, id_completeData)), ]
DEMO = DEMO[,c('RIAGENDR','RIDAGEYR','RIDRETH1')]
\end{CodeInput}
\end{CodeChunk}
\end{comment}

\paragraph{Step 3: Minute‑level step counts (PhysioNet).}  Step counts provide an external instrument for measurement‑error correction.  
We download the minute level step count data from PhysioNet and convert the data into hour level. 
The result is also reshaped into a $n\times24\times7$ array.

\begin{comment}
\begin{CodeChunk}
\begin{CodeInput}
destfile = "nhanes_1440_adeptsteps.csv.xz"
url = paste0("https://physionet.org/files/minute-level-step-count-nhanes/",
              "1.0.1/csv/nhanes_1440_adeptsteps.csv.xz")
download.file(url, destfile,
              method = "curl", extra = "--retry 10 --max-time 600")
steps_data = readr::read_csv(destfile)

steps_data = steps_data[steps_data$SEQN %in% id_completeData,]
steps_data = steps_data[steps_data$PAXDAYM %in% 2:8,]
steps_data = steps_data[order(match(steps_data$SEQN, id_completeData)),]

stepCounts = as.matrix(steps_data[,-(1:3)])
col_groups = ceiling(seq_along(1:1440) / 60)
stepCounts = sapply(unique(col_groups), 
                    function(i) rowSums(stepCounts[, col_groups == i]))
stepCounts = as.data.frame(stepCounts)
colnames(stepCounts) = paste('hr',0:23,sep = '_')
stepCounts = cbind(steps_data[,c("SEQN", "PAXDAYM")],stepCounts)
day_list_SC = split(stepCounts[,paste('hr',0:23,sep = '_')], stepCounts$PAXDAYM)
day_list_SC = lapply(day_list_SC, as.matrix)
SC_array = abind::abind(day_list_SC, along = 3)
\end{CodeInput}
\end{CodeChunk}
\end{comment}

Individuals that have incomplete data or with missing values are removed.  

\begin{comment}
\begin{CodeChunk}
\begin{CodeInput}
have.missing = 
  apply(PA_array, 1, anyNA) | 
  apply(SC_array, 1, anyNA) |
  apply(BMX, 1, anyNA) | 
  apply(DEMO, 1, anyNA)
  PA_array = PA_array[!have.missing,,]
  SC_array = SC_array[!have.missing,,]
  BMX = BMX[!have.missing,,drop = FALSE]
  DEMO = DEMO[!have.missing,,drop = FALSE]
\end{CodeInput}
\end{CodeChunk}
\end{comment}

\begin{figure}[htbp]
  \centering
  \includegraphics[width=0.9\linewidth]{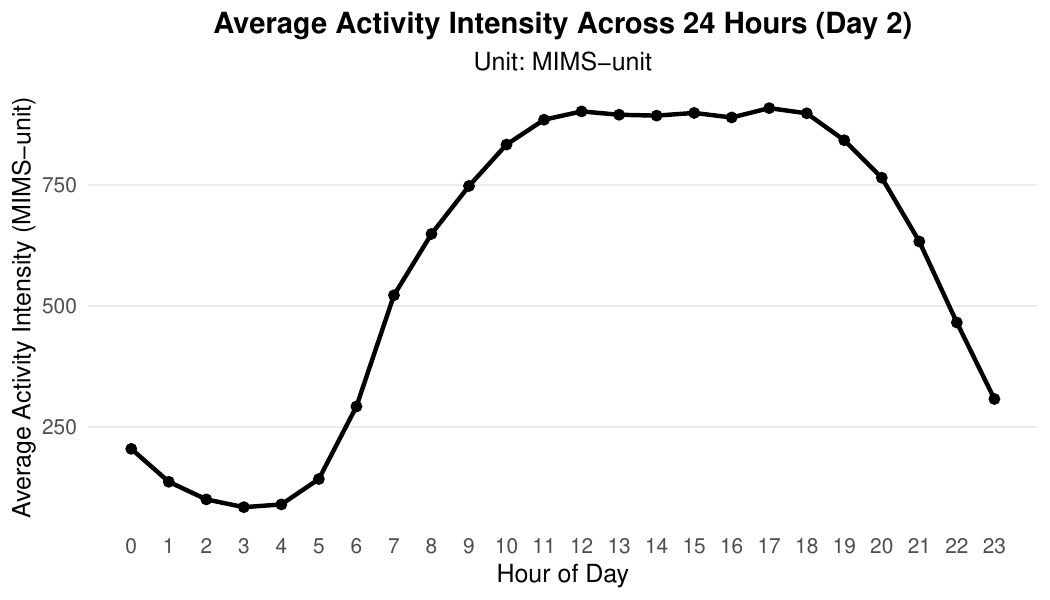}
  \caption{Average activity intensity across 24~hours on the second study day (unit: MIMS-unit).}
  \label{fig:avg-activity-day2}
\end{figure}

\paragraph{Descriptive statistics.} After filtering, the analytic sample
contains $n=6536$ individuals (50.5\% female, mean age $36.8$ years). 
Figure~\ref{fig:avg-activity-day2} shows the mean hourly accelerometer profile of day 2.

\subsection[Data analysis]{Data analysis}

For demonstration, we analyse a random subsample of 100 participants. 
\begin{CodeChunk}
\begin{CodeInput}
set.seed(4)
ind_sub = sample(1:nrow(DEMO), 100, replace = FALSE)
SC_array  = SC_array[ind_sub,,]
PA_array  = PA_array[ind_sub,,]
BMX = BMX[ind_sub,,drop = FALSE]
DEMO = DEMO[ind_sub,,drop = FALSE]
\end{CodeInput}
\end{CodeChunk}

We first fit the naive SoFR and quantile SoFR model for the mean and median functional effects respectively
with the physical activity data of day 2 using the tools provided by \pkg{MECfda}. 

\begin{CodeChunk}
\begin{CodeInput}
library(MECfda)

PA_day2 = PA_array[,,1]

res1 = fcRegression(
  Y = BMX, 
  FC = functional_variable(PA_day2, 0, 24, 0:23), 
  Z = DEMO,
  basis.order = 5, basis.type = c('Bspline'))

res2 = fcQR(
  Y = BMX, 
  FC = functional_variable(PA_day2, 0, 24, 0:23), 
  Z = DEMO,
  basis.order = 5, basis.type = c('Bspline'))
\end{CodeInput}
\end{CodeChunk}

Next, we demonstrate the bias-correction tools in \pkg{MECfda}. 
For the methods that do not require repeated measurements, we only use the data of day 2. 
Please note that the function \fct{fcQR\_CLS} only allows numeric scalar covariates.
We need to convert the matrix/data frame of scalar covariates (argument \code{data.Z}) to actual design matrix before we input it to the function \fct{fcQR\_CLS}. 
The corrected loss score method and SIMEX method need heavy computation load, here we only use the data of a subset of the participants for demonstration. 

\begin{CodeChunk}
\begin{CodeInput}
res3 = ME.fcRegression_MEM(
  data.Y = BMX,
  data.W = PA_array,
  data.Z = DEMO,
  method = 'UP_MEM',
  family.W = "gaussian",
  basis.type = 'Bspline')

SC_day2 = SC_array[,,1]
res4 = ME.fcQR_IV.SIMEX(
  data.Y = BMX,
  data.W = PA_day2,
  data.Z = DEMO,
  data.M = SC_day2,
  t_interval = c(0, 24),
  t_points = 0:23,
  basis.order = 5L,
  bs_degree = 3L,
  tau = 0.5,
  basis.type = 'Bspline')

DEMO_des = model.matrix(~ . , data = DEMO)
DEMO_des = DEMO_des[,-1,drop = FALSE]
res5 = ME.fcQR_CLS(
  data.Y = BMX[1:200,,drop = FALSE],
  data.W = PA_array[1:200,,],
  data.Z = DEMO_des[1:200,],
  t_interval = c(0, 24),
  t_points = 0:23,
  tau = 0.5,
  grid_k = 4:7,
  grid_h = 1:2)
        
res6 = ME.fcLR_IV(
      data.Y = BMX,
      data.W = PA_day2,
      data.M = SC_day2)
\end{CodeInput}
\end{CodeChunk}

\begin{figure}[htbp]
  \centering
  \begin{subfigure}[t]{0.3\textwidth}
    \centering
    \includegraphics[width=\linewidth]{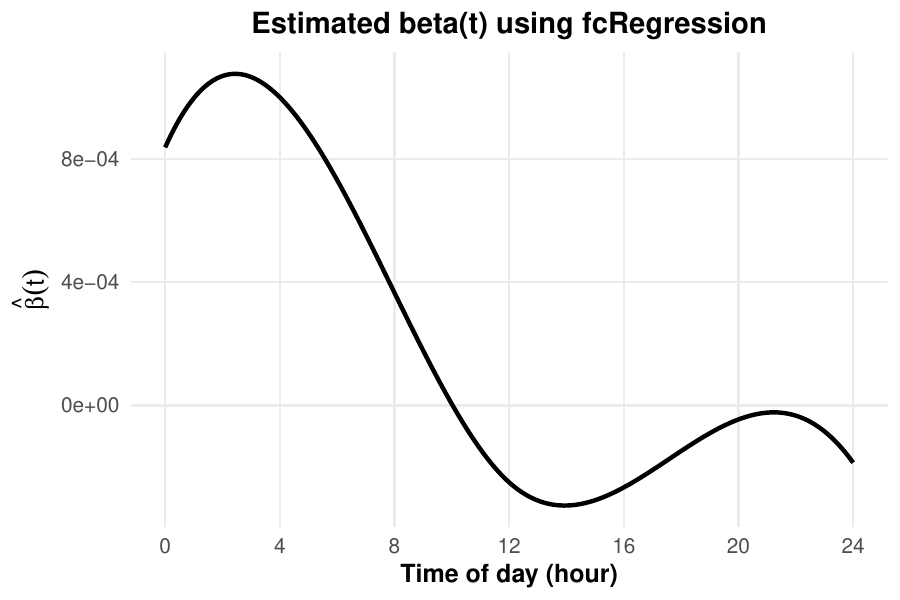}
    \caption{fcRegression}
  \end{subfigure}
  \hfill
  \begin{subfigure}[t]{0.3\textwidth}
    \centering
    \includegraphics[width=\linewidth]{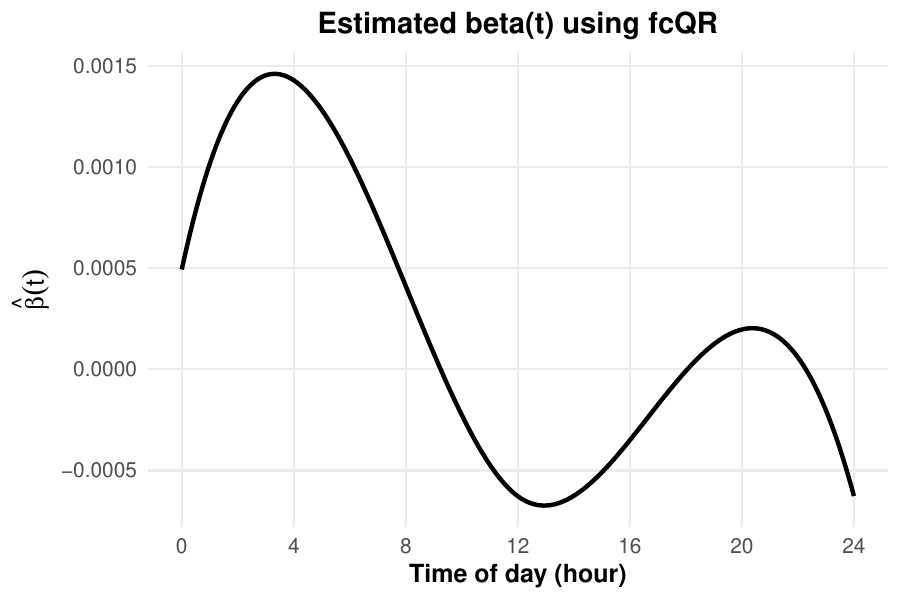}
    \caption{fcQR}
  \end{subfigure}
  \hfill
  \begin{subfigure}[t]{0.3\textwidth}
    \centering
    \includegraphics[width=\linewidth]{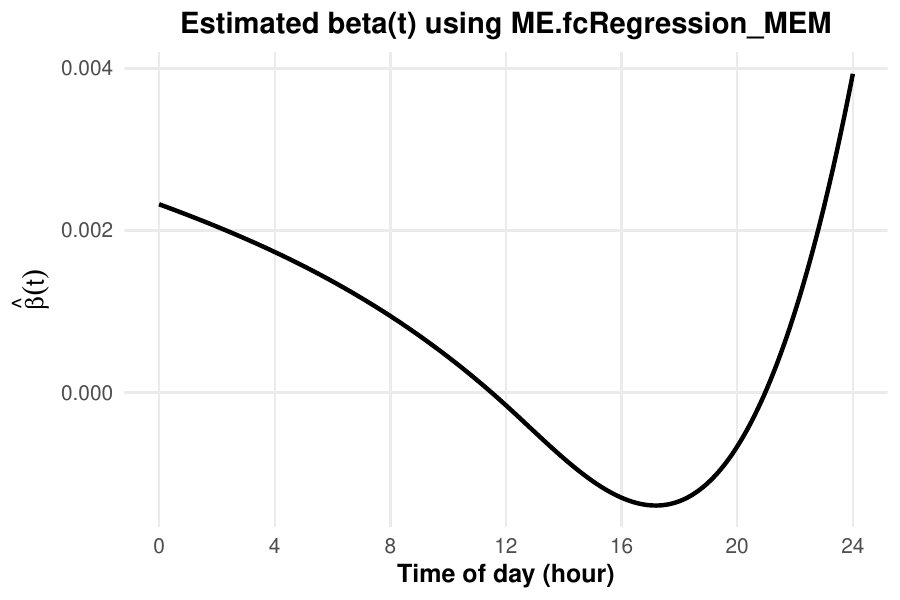}
    \caption{ME.fcRegression\_MEM}
  \end{subfigure}

  \vspace{0.5em}  % space between rows

  \begin{subfigure}[t]{0.3\textwidth}
    \centering
    \includegraphics[width=\linewidth]{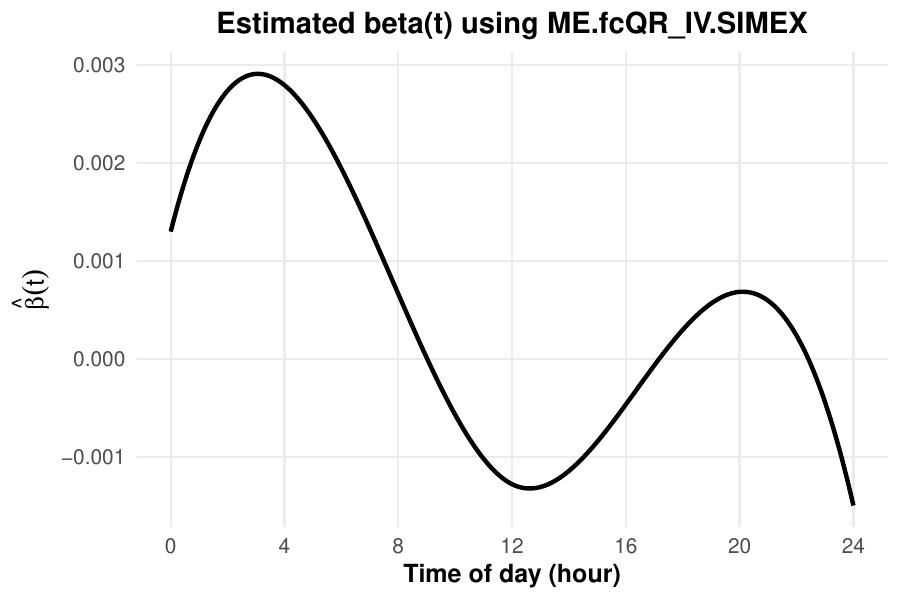}
    \caption{ME.fcQR\_IV.SIMEX}
  \end{subfigure}
  \hfill
  \begin{subfigure}[t]{0.3\textwidth}
    \centering
    \includegraphics[width=\linewidth]{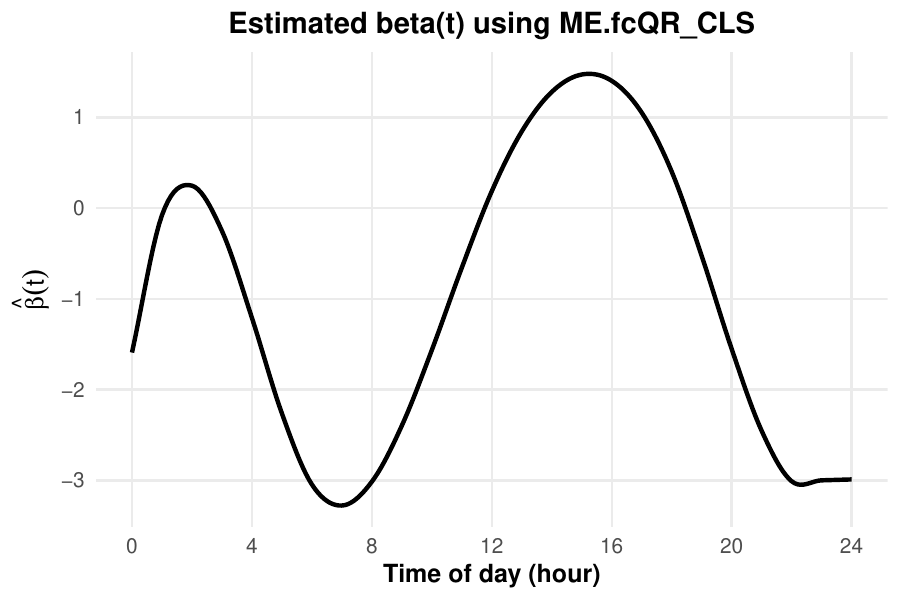}
    \caption{ME.fcQR\_CLS}
  \end{subfigure}
  \hfill
  \begin{subfigure}[t]{0.3\textwidth}
    \centering
    \includegraphics[width=\linewidth]{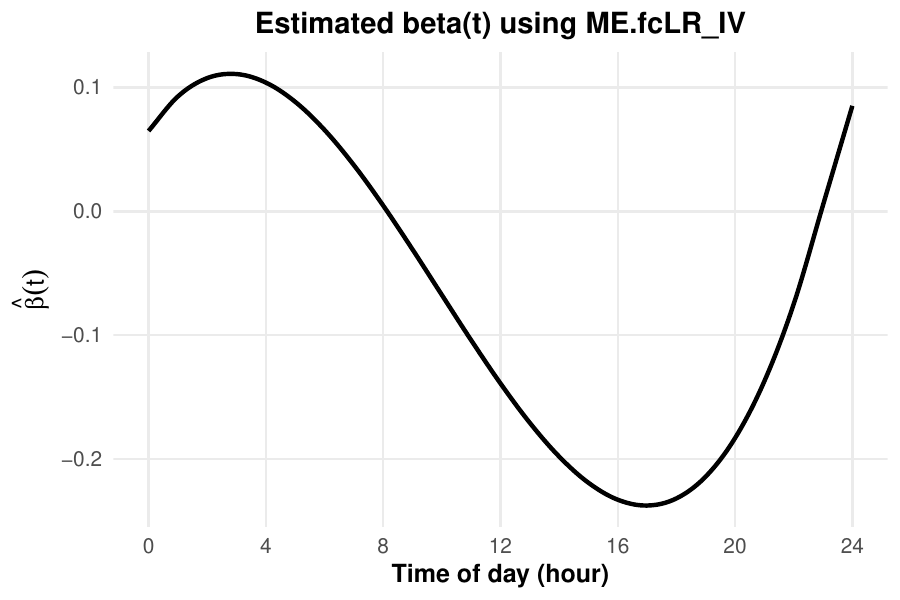}
    \caption{ME.fcLR\_IV}
  \end{subfigure}

  \caption{Estimated coefficient functions $\hat\beta(t)$ obtained from different SoFR methods implemented in \pkg{MECfda}.}
  \label{fig:beta hat}
\end{figure}

\pkg{MECfda} provides built-in visualization tools to help users examine estimation results. However, these built-in methods are relatively rudimentary. Therefore, we have implemented custom functions to visualize the estimated coefficient function $\hat\beta(t)$. 
Figure~\ref{fig:beta hat} presents the estimated $\hat\beta(t)$ obtained using various SoFR tools included in \pkg{MECfda}, applied to a small subset of the data. (Note that we only use a subset here to illustrate the usage of \pkg{MECfda}, rather than analyzing the full dataset.)

\subsection{Discussion}

This case study shows how \pkg{MECfda} streamlines functional measurement‑error correction in real-world epidemiological datasets. 
This example demonstrates the application of the regression and bias-correction routines implemented in the \pkg{MECfda} package. 
In practical applications, users should select model specifications, variables, and associated parameters with meticulous care to ensure robust and precise estimation.

\paragraph{Supplementary materials.}  
The complete, fully annotated, script for reproducing the results and 
Figures are supplied in Appendix~\ref{Example project codes}.

% \section{Computational details}

% \subsection{Numerical Computation of Integrals}

% For integral
% $$\int_{\Omega} \rho_{k}(t) X_{i}(t) dt\ ,$$
% we use
% $$\frac{\mu(\Omega)}{|T|}\sum_{t\in T} \rho_{k}(t) X_{i}(t)$$
% to numerically compute it,
% where $T$ is the measurement (time) points of $X_{i}(t)$, and
% $|T|$ represents the cardinal number of $T$, 
% $\mu(\cdot)$ denotes the Lebesgue measurement of sets. 

% \section*{Acknowledgments}

\bibliography{refs}

\newpage

\begin{appendix}

% \section{More technical details} \label{app:technical}

\section{Basis functions}\label{def basis function}

\paragraph{Fourier basis}

On the interval $[t_0, t_0+T]$, the Fourier basis consists of
$$
\frac{1}{2}, \quad \cos\left(\frac{2\pi}{T}k(x-t_0)\right), \quad \sin\left(\frac{2\pi}{T}k(x-t_0)\right), \quad k = 1, 2, \dots
$$

\paragraph{B-splines basis}

A B-spline basis $\{B_{i,p}(x)\}_{i=-p}^{k}$ on the interval $[t_0, t_{k+1}]$ is defined recursively as follows:

\begin{itemize}
    \item For $p = 0$,
    \begin{equation}
    B_{i,0}(x) =
    \begin{cases}
        I_{(t_i, t_{i+1}]}(x), & i = 0, 1, \dots, k, \\
        0, & \text{otherwise}.
    \end{cases}
    \end{equation}

    \item For $p > 0$, the recursion is given by
    \begin{equation}
    B_{i,p}(x) = \frac{x-t_i}{t_{i+p}-t_i} B_{i,p-1}(x) + \frac{t_{i+p+1}-x}{t_{i+p+1}-t_{i+1}} B_{i+1,p-1}(x).
    \end{equation}
\end{itemize}

At discontinuities in the interval $(t_0,t_k)$, 
the basis function is defined as its limit,
\begin{equation}
B_{i,p}(x) = \lim_{t\to x} B_{i,p}(t),
\end{equation}
to ensure continuity.

\paragraph{Eigenfunction basis}

Suppose $K(s,t)\in L^2(\Omega\times \Omega)$ and 
$f(t)\in L^2(\Omega)$. 
Then $K$ induces a linear operator $\mathcal{K}$ defined by 
\begin{equation}
(\mathcal{K}f)(x) = \int_{\Omega} K(t,x) f(t) dt.
\end{equation}

If $\xi \in L^2(\Omega)$ satisfies 
\begin{equation}
\mathcal{K}\xi = \lambda \xi,
\end{equation}
where $\lambda\in \mathbb{C}$,
we call $\xi$ an eigenfunction (or eigenvector) of 
$\mathcal{K}$, and $\lambda$ an eigenvalue associated with $\xi$. 

For a stochastic process $\{X(t), t\in\Omega\}$,
the orthogonal basis $\{\xi_k\}_{k=1}^\infty$
corresponding to eigenvalues $\{\lambda_k\}_{k=1}^\infty$ 
with $\lambda_1\geq\lambda_2\geq\dots$, 
is induced by 
\begin{equation}
K(s,t) = \operatorname{Cov}(X(t),X(s)),
\end{equation}
and is called a functional principal component (FPC) basis for $L^2(\Omega)$.

The eigenfunction basis depends on the covariance function $K(s,t)$. 
In most cases, we cannot analytically express the basis functions in the eigenfunction basis. 
Instead, the covariance function is estimated from data, and the eigenfunctions are computed numerically.

\section{Software reference}\label{Software reference}

The package \pkg{MECfda} provides tools to perform scalar-on-function regression 
and bias correction methods due to measurement error. 

\subsection*{S4 class \class{functional\_variable}}

In the \pkg{MECfda} package, the S4 class \class{functional\_variable} represents functional data stored in matrix form. The main slots include:

\begin{itemize}
    \item \textbf{\code{X}}: The data matrix $(x_{ij})$;
    \item \textbf{\code{t\_0}} and \textbf{\code{period}}: The left endpoint and length of the function’s domain, respectively (the class only supports domains in interval format);
    \item \textbf{\code{t\_points}}: A vector of the time points at which the measurements are recorded.
\end{itemize}

A dedicated method, \fct{dim}, returns the number of subjects and the number of measurement points for a \class{functional\_variable} object.

% \textbf{Example codes}:
% \begin{Code}
%     fv = functional_variable(
%       X = matrix(rnorm(10*24),10,24),
%       t_0 = 0,
%       period = 1,
%       t_points = (0:9)/10
%     )
%     dim(fv)
% \end{Code}

\subsection*{S4 class \class{Fourier\_series}}

In the \pkg{MECfda} package, the S4 class \class{Fourier\_series} represents a Fourier series of the form:
\begin{equation}
\frac{a_0}{2} + \sum_{k=1}^{p_a} a_k \cos\left(\frac{2\pi k (x-t_0)}{T}\right) + \sum_{k=1}^{p_b} b_k \sin\left(\frac{2\pi k (x-t_0)}{T}\right),\quad x \in [t_0, t_0+T].
\end{equation}

Its main slots include:
\begin{itemize}
    \item \textbf{\code{double\_constant}}: The value of $a_0$;
    \item \textbf{\code{cos}} and \textbf{\code{sin}}: The coefficients $a_k$ and $b_k$ for the cosine and sine terms, respectively;
    \item \textbf{\code{k\_cos}} and \textbf{\code{k\_sin}}: The frequency values corresponding to the cosine and sine coefficients;
    \item \textbf{\code{t\_0}} and \textbf{\code{period}}: The left endpoint, $t_0$, and length, $T$, of the domain (interval $[t_0, t_0+T]$), respectively.
\end{itemize}

Additional methods such as \fct{plot}, \fct{FourierSeries2fun}, and \fct{extractCoef} are provided for visualization, function evaluation, and coefficient extraction.

% \textbf{Example codes}:
% \begin{Code}
%     fsc = Fourier_series(
%       double_constant = 3,
%       cos = c(0,2/3),
%       sin = c(1,7/5),
%       k_cos = 1:2,
%       k_sin = 1:2,
%       t_0 = 0,
%       period = 1
%     )
%     plot(fsc)
%     FourierSeries2fun(fsc,seq(0,1,0.05))
%     extractCoef(fsc)
% \end{Code}

The object \code{fsc} represents the summation
$$\frac32 + \frac23 \cos(2\pi\cdot2x) + \sin(2\pi x) + \frac75\sin(2\pi\cdot2x).$$

\subsection*{S4 class \class{bspline\_basis} and \class{bspline\_series}}

The S4 class \class{bspline\_basis} represents a B-spline basis, $\{B_{i,p}(x)\}_{i=-p}^{k}$, on the interval $[t_0, t_{k+1}]$. Its key slots include:

\begin{itemize}
    \item \textbf{\code{Boundary.knots}}: The boundary $[t_0, t_{k+1}]$ (default is $[0,1]$);
    \item \textbf{\code{knots}}: The internal spline knots, $(t_1,\dots,t_k)$ (automatically generated as equally spaced, $t_j = t_0 + j\cdot\frac{t_{k+1}-t_0}{k+1}$, if not assigned);
    \item \textbf{\code{intercept}}: Indicates whether an intercept is included (default is \code{TRUE});
    \item \textbf{\code{df}}: Degrees of freedom of the basis, which is the number of splines, equal to $p + k + 1$ (by default, $k = 0$ and $\text{df} = p+1$);
    \item \textbf{\code{degree}}: The degree of the spline, which is the degree of piecewise polynomials, $p$ (default is 3).
\end{itemize}

The S4 class \class{bspline\_series} represents the linear combination:
\begin{equation}
\sum_{i=0}^{k} b_i B_{i,p}(x),
\end{equation}
where the slot \textbf{\code{coef}} stores the coefficients $b_i$ ($i = 0,\dots,k$) and \textbf{\code{bspline\_basis}} specifies the associated basis (represented by a \class{bspline\_basis} object).

Methods such as \fct{plot} and \fct{bsplineSeries2fun} are provided for visualization and function evaluation.

% \textbf{Example codes}:
% \begin{Code}
%     bsb = bspline_basis(
%       Boundary.knots = c(0,24),
%       df             = 7,
%       degree         = 3
%     )
%     bss = bspline_series(
%       coef = c(2,1,3/4,2/3,7/8,5/2,19/10),
%       bspline_basis = bsb
%     )
%     plot(bss)
%     bsplineSeries2fun(bss,seq(0,24,0.5))
% \end{Code}

The object \code{bsb} represents 
$\{B_{i,3}(x)\}_{i=-3}^{0}$, 
and the object \code{bss} represents  
$$2B_{i,-3}(x)+B_{i,-2}(x)+\frac34B_{i,-1}(x)+\frac23B_{i,0}(x) + \frac78B_{i,1}(x) + 
\frac52B_{i,2}(x) +\frac{19}{10}B_{i,3}(x),$$
where $x\in[t_0,t_4]$ and $t_0=0$, $t_k = t_{k-1}+6$ ,$k=1,2,3,4$. 

\subsection*{S4 Classes \class{numeric\_basis} and \class{numericBasis\_series}}

In the \pkg{MECfda} package, the S4 class \class{numeric\_basis} is designed to numerically represent a finite sequence of functions $\{\rho_k\}_{k=1}^p$ by their values at a finite set of points within their domain. In this context, all functions share the same domain, which is assumed to be an interval. 
The key slots include:

\begin{itemize}
    \item \textbf{\code{basis\_function}}: The matrix $(\zeta_{jk})_{m \times p}$, where each element is defined as $\zeta_{jk} = \rho_k(t_j)$ for $j = 1, \dots, m$ and $k = 1, \dots, p$. In this matrix, each row corresponds to a point in the domain, and each column corresponds to a specific basis function.
    \item \textbf{\code{t\_points}}: A numeric vector representing the points in the domain at which the basis functions are evaluated. The $j$-th element of this vector corresponds to the $j$-th row of the \code{basis\_function} matrix.
    \item \textbf{\code{t\_0}} and \textbf{\code{period}}: The left endpoint and length of the domain, respectively.
\end{itemize}

Additionally, the package provides an S4 class \class{numericBasis\_series}, which represents a linear combination of the basis functions represented by a \class{numeric\_basis} object. 
The key slots include:

\begin{itemize}
    \item \textbf{\code{coef}}: The linear coefficients for the summation series.
    \item \textbf{\code{numeric\_basis}}: Function basis as represented by a \class{numeric\_basis} object.
\end{itemize}

\subsection*{Function \fct{basis2fun}}

The generic function \fct{basis2fun} is provided for \class{Fourier\_series}, \class{bspline\_series}, and \class{numericBasis\_series} objects. 
\begin{itemize}
    \item For a \class{Fourier\_series} object, it is equivalent to \fct{FourierSeries2fun}.
    \item For a \class{bspline\_series} object, it is equivalent to \fct{bsplineSeries2fun}.
    \item For a \class{numericBasis\_series} object, it is equivalent to \fct{numericBasisSeries2fun}.
\end{itemize}

% \textbf{Example codes}:
% \begin{Code}
%     basis2fun(fsc,seq(0,1,0.05))
%     basis2fun(bss,seq(0,24,0.5))
% \end{Code}

\subsection*{Basis Expansion Methods for the \class{functional\_variable} class objects}

The \pkg{MECfda} package provides the methods 
\fct{fourier\_basis\_expansion}, \\\fct{bspline\_basis\_expansion}, \fct{FPC\_basis\_expansion}, and \fct{numeric\_basis\_expansion}
for the \class{functional\_variable} class, 
which perform basis expansion using the Fourier basis, B-spline basis, functional principal component (FPC) basis, 
and numerical basis, respectively.

% \textbf{Example codes}:
% \begin{CodeChunk}
% \begin{CodeInput}
%     data(MECfda.data.sim.0.0)
%     fv = MECfda.data.sim.0.0$FC[[1]]
%     BE.fs = fourier_basis_expansion(fv,5L)
%     BE.bs = bspline_basis_expansion(fv,5L,3L)
    
%     n<-50; ti<-seq(0,1,length.out=101)
%     X<-t(sin(2*pi*ti)%*%t(rnorm(n,0,1)))
%     object = functional_variable(X = X, t_0 = 0, period = 1, t_points = ti)
%     BE.fpc = FPC_basis_expansion(object,3L)
    
%     nb = splines::bs(
%       fv@t_points,
%       df = 5L,
%       degree = 3L,
%       intercept = TRUE,
%       Boundary.knots = c(fv@t_0, fv@t_0 + fv@period))
%     nb = numeric_basis(
%       basis_function = unclass(nb),
%       t_points = fv@t_points,
%       t_0      = fv@t_0,
%       period   = fv@period)
%     BE.num = numeric_basis_expansion(fv,nb)
% \end{CodeInput}
% \end{CodeChunk}

\subsection*{Function \fct{fcRegression}}

\begin{CodeChunk}
\begin{CodeInput}
fcRegression(Y, FC, Z, formula.Z, family = gaussian(link = "identity"),
         basis.type = c("Fourier", "Bspline"), basis.order = 6L, 
         bs_degree = 3)
\end{CodeInput}
\end{CodeChunk}

The \pkg{MECfda} package provides the function \fct{fcRegression} for fitting generalized linear mixed-effects models—including ordinary linear models and generalized linear models with fixed and random effects—by discretizing functional covariates using basis expansion. The function can solve a linear model of the form:
\begin{equation}
g\bigl(E(Y_i\mid X_i,Z_i)\bigr) = \sum_{l=1}^{L} \int_{\Omega_l} \beta_l(t) X_{li}(t) \, dt + (1,Z_i^T)\gamma.
\end{equation}

It supports one or multiple functional covariates as fixed effects and zero, one, or multiple scalar covariates, which can be specified as either fixed or random effects. The response variable, functional covariates, and scalar covariates are supplied separately via the arguments \code{Y}, \code{FC}, and \code{Z}, respectively, allowing for great flexibility in data format.

\paragraph{Response variable:}  
The response can be provided as an atomic vector, a one-column matrix, or a data frame. However, a one-column data frame or matrix with a column name is recommended so that the response variable's name is clearly specified.

\paragraph{Functional covariates:}  
The functional covariates can be input as a \class{functional\_variable} object, a matrix, a data frame, or a list of these objects. If a single \class{functional\_variable} object (or matrix or data frame) is provided via \code{FC}, the model includes one functional covariate. If a list is provided, the model accommodates multiple functional covariates, with each element in the list representing a distinct covariate.

\paragraph{Scalar covariates:}  
Scalar covariates can be provided as a matrix, data frame, atomic vector, or even \code{NULL} if no scalar covariates are present. If \code{Z} is omitted or set to \code{NULL}, no scalar covariate is included, and the argument \code{formula.Z} should also be \code{NULL} or omitted. When an atomic vector is used for \code{Z}, it represents a single scalar covariate (without a name). Hence, even when only one scalar covariate is included, it is recommended to supply it as a matrix or data frame with a column name. The \code{formula.Z} argument specifies which parts of \code{Z} to use and whether scalar covariates should be treated as fixed or random effects.

\paragraph{Key arguments:}
\begin{itemize}
    \item \textbf{\code{family}:} Specifies the response variable's distribution and the link function to be used.
    \item \textbf{\code{basis.type}:} Indicates the type of basis function for the expansion. Options include \code{"Fourier"}, \code{"Bspline"}, and \code{"FPC"}, corresponding to the Fourier basis, B-spline basis, and functional principal component (FPC) basis, respectively.
    \item \textbf{\code{basis.order}:} Specifies the number of basis functions to use. For the Fourier basis (i.e., $\frac{1}{2},\ \sin(kt),\ \cos(kt)$ for $k = 1, \dots, p_f$), \code{basis.order} equals $p_f$. For the B-spline basis $\{B_{i,p}(x)\}_{i=-p}^{k}$, \code{basis.order} is set to $k+p+1$. For the FPC basis, \code{basis.order} represents the number of functional principal components.
    \item \textbf{\code{bs\_degree}:} Specifies the degree of the piecewise polynomials for the B-spline basis; this argument is only necessary when using the B-spline basis.
\end{itemize}

\paragraph{Returned object:}  
The function returns an S3 object of class \class{fcRegression}, which is a list containing the following elements:

\begin{enumerate}
    \item \textbf{\code{regression\_result}:} The result of the regression analysis.
    \item \textbf{\code{FC.BasisCoefficient}:} A list of \class{Fourier\_series}, \class{bspline\_series}, or \class{numericBasis\_series} objects representing the functional linear coefficients of the covariates.
    \item \textbf{\code{function.basis.type}:} The type of basis used.
    \item \textbf{\code{basis.order}:} The number of basis functions used, as specified in the input.
    \item \textbf{\code{data}:} The original data provided to the model.
    \item \textbf{\code{bs\_degree}:} The degree of the B-spline basis, same as specified in the input (included only when the B-spline basis is used).
\end{enumerate}

Additionally, the \fct{predict} method can be used to obtain predicted values from the fitted model, and the \fct{fc.beta} method is available to evaluate the estimated functional coefficient parameters $\beta_l(t)$.

% \textbf{Example codes}:
% \begin{CodeChunk}
% \begin{CodeInput}
%     data(MECfda.data.sim.0.0)
%     res = fcRegression(FC = MECfda.data.sim.0.0$FC, 
%                        Y=MECfda.data.sim.0.0$Y, 
%                        Z=MECfda.data.sim.0.0$Z,
%                        family = gaussian(link = "identity"),
%                        basis.order = 5, basis.type = c('Bspline'),
%                        formula.Z = ~ Z_1 + (1|Z_2))
%     t = (0:100)/100
%     plot(x = t, y = fc.beta(res,1,t), ylab = expression(beta[1](t)))
%     plot(x = t, y = fc.beta(res,2,t), ylab = expression(beta[2](t)))
%     data(MECfda.data.sim.1.0)
%     predict(object = res, newData.FC = MECfda.data.sim.1.0$FC,
%             newData.Z = MECfda.data.sim.1.0$Z)
% \end{CodeInput}
% \end{CodeChunk}

\subsection*{Function \fct{fcQR}}

\begin{CodeChunk}
\begin{CodeInput}
fcQR(Y, FC, Z, formula.Z, tau = 0.5, basis.type = c("Fourier", "Bspline"),
     basis.order = 6L, bs_degree = 3)
\end{CodeInput}
\end{CodeChunk}

The \pkg{MECfda} package provides the function \fct{fcQR} for fitting quantile linear regression models by discretizing functional covariates using basis expansion. The function can solve a linear model of the form:
\begin{equation}
Q_{Y_i\mid X_i,Z_i}(\tau) = \sum_{l=1}^L\int_{\Omega_l} \beta_l(\tau,t) X_{li}(t) dt +  (1,Z_i^T)\gamma.
\end{equation}

The function supports one or multiple functional covariates, as well as zero, one, or multiple scalar-valued covariates. Data input follows the same guidelines as for the \fct{fcRegression} function, and the treatment of scalar covariates is determined by the \code{formula.Z} argument. The primary difference between \fct{fcQR} and \fct{fcRegression} is that the quantile linear regression model does not incorporate random effects. The quantile $\tau$ is specified by the argument \code{tau}, and the type and parameters of the basis functions are defined by the \code{basis.type}, \code{basis.order}, and \code{bs\_degree} arguments, just as in \fct{fcRegression}.

\paragraph{Returned object:}  
The function returns an S3 object of class \class{fcQR}, which is a list containing the following elements:

\begin{enumerate}
    \item \textbf{\code{regression\_result}:} The result of the regression analysis.
    \item \textbf{\code{FC.BasisCoefficient}:} A list of \class{Fourier\_series}, \class{bspline\_series}, or \class{numericBasis\_series} objects representing the functional linear coefficients of the covariates.
    \item \textbf{\code{function.basis.type}:} The type of basis used.
    \item \textbf{\code{basis.order}:} The number of basis functions used, as specified in the input.
    \item \textbf{\code{data}:} The original data provided to the model.
    \item \textbf{\code{bs\_degree}:} The degree of the B-spline basis, same as specified in the input (included only when the B-spline basis is used).
\end{enumerate}

Additionally, the \fct{predict} method can be used to obtain predicted values from the fitted model, and the \fct{fc.beta} method is available to evaluate the estimated functional coefficient parameters $\beta_l(t)$.

% \textbf{Example codes}:
% \begin{CodeChunk}
% \begin{CodeInput}
% data(MECfda.data.sim.0.0)
% res = fcQR(FC = MECfda.data.sim.0.0$FC, 
%            Y=MECfda.data.sim.0.0$Y, 
%            Z=MECfda.data.sim.0.0$Z,
%            tau = 0.5,
%            basis.order = 5, basis.type = c('Bspline'),
%            formula.Z = ~ .)
% t = (0:100)/100
% plot(x = t, y = fc.beta(res,1,t), ylab = expression(beta[1](t)))
% plot(x = t, y = fc.beta(res,2,t), ylab = expression(beta[2](t)))
% data(MECfda.data.sim.1.0)
% predict(object = res, newData.FC = MECfda.data.sim.1.0$FC,
%         newData.Z = MECfda.data.sim.1.0$Z)
% \end{CodeInput}
% \end{CodeChunk}

\subsection*{Function \fct{ME.fcRegression\_MEM}}

\begin{CodeChunk}
\begin{CodeInput}
ME.fcRegression_MEM(
  data.Y,
  data.W,
  data.Z,
  method = c("UP_MEM", "MP_MEM", "average"),
  t_interval = c(0, 1),
  t_points = NULL,
  d = 3,
  family.W = c("gaussian", "poisson"),
  family.Y = "gaussian",
  formula.Z,
  basis.type = c("Fourier", "Bspline"),
  basis.order = NULL,
  bs_degree = 3,
  smooth = FALSE,
  silent = TRUE
)

\end{CodeInput}
\end{CodeChunk}

The \pkg{MECfda} package provides the function \fct{ME.fcRegression\_MEM} for applying bias-correction estimation methods. In this function, the response variable, functional covariates, and scalar covariates are supplied separately via the arguments \code{data.Y}, \code{data.W}, and \code{data.Z}, respectively.

\paragraph{Input data:}
\begin{itemize}
    \item \textbf{\code{data.Y} (Response Variable):}  
    The response variable can be provided as an atomic vector, a one-column matrix, or a data frame. However, a one-column data frame with a column name is recommended so that the response variable is explicitly identified.

    \item \textbf{\code{data.W} (Measurement Data):}  
    This argument represents $W$, the observed measurement of the true functional covariate $X$ in the statistical model. It should be provided as a three-dimensional array where each row corresponds to a subject, each column to a measurement (time) point, and each layer to a separate observation.

    \item \textbf{\code{data.Z} (Scalar Covariates):}  
    The scalar covariates can be input as a matrix, data frame, or atomic vector. If the model does not include any scalar covariates, \code{data.Z} can be omitted or set to \code{NULL}. For a single scalar covariate, although an atomic vector is acceptable, a data frame or matrix with a column name is recommended. For multiple scalar covariates, use a matrix or data frame with named columns.
\end{itemize}

\paragraph{Key arguments:}
\begin{itemize}
    \item \textbf{\code{method}:} Specifies the method for constructing the substitute for $X$; available options are \code{"UP\_MEM"}, \code{"MP\_MEM"}, and \code{"average"}.
    \item \textbf{\code{t\_interval}:} Defines the domain of the functional covariate as a two-element vector (default is \code{c(0,1)}, representing the interval $[0,1]$).
    \item \textbf{\code{t\_points}:} Specifies the sequence of measurement time points (default is \code{NULL}).
    \item \textbf{\code{d}:} When using the \code{"MP\_MEM"} method, this argument sets the number of measurement points involved (default is 3, which is also the minimum value).
    \item \textbf{\code{family.W}:} Specifies the distribution of $W$ given $X$; available options are \code{"gaussian"} and \code{"poisson"}.
    \item \textbf{\code{family.Y}:} Describes the error distribution and link function for the response variable (see \code{stats::family} for details).
    \item \textbf{\code{formula.Z}:} Indicates which parts of \code{data.Z} to include in the model and whether to treat the scalar covariates as fixed or random effects.
    \item \textbf{\code{basis.type}:} Indicates the type of basis function for the expansion. Options include \code{"Fourier"}, \code{"Bspline"}, and \code{"FPC"}, corresponding to the Fourier basis, B-spline basis, and functional principal component basis, respectively.
    \item \textbf{\code{basis.order}:} Specifies the number of basis functions to use. For the Fourier basis (i.e., $\frac{1}{2},\ \sin(kt),\ \cos(kt)$ for $k = 1, \dots, p_f$), \code{basis.order} equals $p_f$. For the B-spline basis $\{B_{i,p}(x)\}_{i=-p}^{k}$, \code{basis.order} is set to $k+p+1$. For the FPC basis, \code{basis.order} represents the number of functional principal components.
    \item \textbf{\code{bs\_degree}:} Specifies the degree of the piecewise polynomials for the B-spline basis; this argument is only necessary when using the B-spline basis.
    \item \textbf{\code{smooth}:} Specifies whether to smooth the substitution for $X$ (default is \code{FALSE}).
    \item \textbf{\code{silent}:} Controls whether the function displays its progress during execution (default is \code{TRUE}).
\end{itemize}

\paragraph{Returned object:}  
The function \fct{ME.fcRegression\_MEM} returns an object of class \class{fcRegression}.

\paragraph{Additional function:}  
The package also provides the function \fct{MEM\_X\_hat}, which returns the estimated $\hat{X}(t)$ in this bias-correction method.

% \textbf{Example codes}:
% \begin{CodeChunk}
% \begin{CodeInput}
% data(MECfda.data.sim.0.1)
% res = ME.fcRegression_MEM(data.Y = MECfda.data.sim.0.1$Y,
%                           data.W = MECfda.data.sim.0.1$W,
%                           data.Z = MECfda.data.sim.0.1$Z,
%                           method = 'UP_MEM',
%                           family.W = "gaussian",
%                           basis.type = 'Bspline')
% \end{CodeInput}
% \end{CodeChunk}

\subsection*{Function \fct{ME.fcQR\_IV.SIMEX}}

\begin{CodeChunk}
\begin{CodeInput}
ME.fcQR_IV.SIMEX(
  data.Y,
  data.W,
  data.Z,
  data.M,
  tau = 0.5,
  t_interval = c(0, 1),
  t_points = NULL,
  formula.Z,
  basis.type = c("Fourier", "Bspline"),
  basis.order = NULL,
  bs_degree = 3
)
\end{CodeInput}
\end{CodeChunk}

The \pkg{MECfda} package provides the function \fct{ME.fcRegression\_MEM} for applying bias-correction estimation methods. In this function, the response variable, functional covariates, and scalar covariates are supplied separately via the arguments \code{data.Y}, \code{data.W}, and \code{data.Z}, respectively.

\paragraph{Input data:}
\begin{itemize}
    \item \textbf{\code{data.Y} (Response Variable):}  
    The response variable can be provided as an atomic vector, a one-column matrix, or a data frame. However, a one-column data frame with a column name is recommended so that the response variable is explicitly identified.

    \item \textbf{\code{data.W} (Measurement Data):}  
    This argument represents $W$, the observed measurement of the true functional covariate $X$ in the statistical model. It should be provided as a three-dimensional array where each row corresponds to a subject, each column to a measurement (time) point, and each layer to a separate observation.

    \item \textbf{\code{data.Z} (Scalar Covariates):}  
    The scalar covariates can be input as a matrix, data frame, or atomic vector. If the model does not include any scalar covariates, \code{data.Z} can be omitted or set to \code{NULL}. For a single scalar covariate, although an atomic vector is acceptable, a data frame or matrix with a column name is recommended. For multiple scalar covariates, use a matrix or data frame with named columns.
\end{itemize}

\paragraph{Key arguments:}
\begin{itemize}
    \item \textbf{\code{method}:} Specifies the method for constructing the substitute for $X$; available options are \code{"UP\_MEM"}, \code{"MP\_MEM"}, and \code{"average"}.
    \item \textbf{\code{t\_interval}:} Defines the domain of the functional covariate as a two-element vector (default is \code{c(0,1)}, representing the interval $[0,1]$).
    \item \textbf{\code{t\_points}:} Specifies the sequence of measurement time points (default is \code{NULL}).
    \item \textbf{\code{d}:} When using the \code{"MP\_MEM"} method, this argument sets the number of measurement points involved (default is 3, which is also the minimum value).
    \item \textbf{\code{family.W}:} Specifies the distribution of $W$ given $X$; available options are \code{"gaussian"} and \code{"poisson"}.
    \item \textbf{\code{family.Y}:} Describes the error distribution and link function for the response variable (see \code{stats::family} for details).
    \item \textbf{\code{formula.Z}:} Indicates which parts of \code{data.Z} to include in the model and whether to treat the scalar covariates as fixed or random effects.
    \item \textbf{\code{basis.type}:} Indicates the type of basis function for the expansion. Options include \code{"Fourier"}, \code{"Bspline"}, and \code{"FPC"}, corresponding to the Fourier basis, B-spline basis, and functional principal component basis, respectively.
    \item \textbf{\code{basis.order}:} Specifies the number of basis functions to use. For the Fourier basis (i.e., $\frac{1}{2},\ \sin(kt),\ \cos(kt)$ for $k = 1, \dots, p_f$), \code{basis.order} equals $p_f$. For the B-spline basis $\{B_{i,p}(x)\}_{i=-p}^{k}$, \code{basis.order} is set to $k+p+1$. For the FPC basis, \code{basis.order} represents the number of functional principal components.
    \item \textbf{\code{bs\_degree}:} Specifies the degree of the piecewise polynomials for the B-spline basis; this argument is only necessary when using the B-spline basis.
    \item \textbf{\code{smooth}:} Specifies whether to smooth the substitution for $X$ (default is \code{FALSE}).
    \item \textbf{\code{silent}:} Controls whether the function displays its progress during execution (default is \code{TRUE}).
\end{itemize}

\paragraph{Returned object:}  
The function \texttt{ME.fcQR\_IV.SIMEX} returns an object of class \texttt{ME.fcQR\_IV.SIMEX}, which is a list containing the following elements:

\begin{enumerate}
  \item \texttt{coef.X}: A \texttt{Fourier\_series} or \texttt{bspline\_series} object representing the function-valued coefficient parameter of the function-valued covariate.
  \item \texttt{coef.Z}: The estimated linear coefficients of the scalar covariates.
  \item \texttt{coef.all}: The original estimate of the linear coefficients.
  \item \texttt{function.basis.type}: The type of function basis used.
  \item \texttt{basis.order}: The number of basis functions used (as specified in the input).
  \item \texttt{t\_interval}: A two-element vector representing the interval, which defines the domain of the function-valued covariate.
  \item \texttt{t\_points}: The sequence of measurement (time) points.
  \item \texttt{formula}: The regression model.
  \item \texttt{formula.Z}: A formula object containing only the scalar covariates.
  \item \texttt{zlevels}: The levels of any categorical (non-continuous) scalar covariates.
\end{enumerate}

% \textbf{Example codes}:
% \begin{CodeChunk}
% \begin{CodeInput}
%     rm(list = ls())
%     data(MECfda.data.sim.0.2)
%     res = ME.fcQR_IV.SIMEX(data.Y = MECfda.data.sim.0.2$Y,
%                            data.W = MECfda.data.sim.0.2$W,
%                            data.Z = MECfda.data.sim.0.2$Z,
%                            data.M = MECfda.data.sim.0.2$M,
%                            tau = 0.5,
%                            basis.type = 'Bspline')
% \end{CodeInput}
% \end{CodeChunk}

\subsection*{Function \fct{ME.fcQR\_CLS}}

% \begin{CodeChunk}
% \begin{CodeInput}
% ME.fcQR_CLS(
%   data.Y,
%   data.W,
%   data.Z,
%   tau = 0.5,
%   t_interval = c(0, 1),
%   t_points = NULL,
%   grid_k,
%   grid_h,
%   degree = 45,
%   observed_X = NULL
% )
% \end{CodeInput}
% \end{CodeChunk}

The \pkg{MECfda} package provides the function \fct{ME.fcQR\_IV.SIMEX} for applying its bias-correction estimation method in scalar-on-function quantile regression. 

\paragraph{Input data:}
\begin{description}
  \item[\texttt{data.Y} (Response Variable):] 
  The response variable can be provided as an atomic vector, a one-column matrix, or a data frame. A one-column data frame with a column name is recommended for clarity.

  \item[\texttt{data.W} (Measurement Data):]
  This argument corresponds to $W$, the observed measurement of $X$ in the statistical model. It should be provided as a data frame or matrix where each row represents a subject and each column corresponds to a measurement (time) point.

  \item[\texttt{data.Z} (Scalar Covariates):]
  The scalar covariate data can be omitted (or set to \texttt{NULL}) if no scalar covariates are included in the model. Alternatively, it can be provided as an atomic vector (for a single scalar covariate) or as a matrix/data frame (for multiple scalar covariates). A data frame with column names is recommended.

  \item[\texttt{data.M} (Instrumental Variable):]
  This argument provides the data for $M$, the instrumental variable. It should be supplied as a data frame or matrix where each row represents a subject and each column corresponds to a measurement (time) point.

\end{description}

\paragraph{Key arguments:}
\begin{description}
  \item[\texttt{tau} (Quantile Level):]
  Specifies the quantile level $\tau \in (0,1)$ for the quantile regression model. The default is 0.5.

  \item[\texttt{t\_interval} (Domain of the Function-Valued Covariate):]
  A two-element vector defining the interval over which the function-valued covariate is defined (default is \texttt{c(0,1)}, representing the interval $[0,1]$).

  \item[\texttt{t\_points} (Measurement Time Points):]
  Specifies the sequence of time points at which measurements are taken. The default is \texttt{NULL}.

  \item[\texttt{formula.Z} (Scalar Covariate Formula):]
  This argument determines which components of \texttt{data.Z} are included in the model and how the scalar covariates are treated (i.e., as fixed or random effects).

  \item[\texttt{basis.type}:] 
  Indicates the type of basis function for the expansion. Options include \texttt{'Fourier'}, \texttt{'Bspline'}, and \texttt{'FPC'}, corresponding to the Fourier basis, B-spline basis, and functional principal component basis, respectively.

\item[\texttt{basis.order}:]
  Specifies the number of basis functions to use. For the Fourier basis (i.e., $\frac{1}{2},\ \sin(kt),\ \cos(kt)$ for $k = 1, \dots, p_f$), \texttt{basis.order} equals $p_f$. For the B-spline basis $\{B_{i,p}(x)\}_{i=-p}^{k}$, \texttt{basis.order} is set to $k + p + 1$. For the FPC basis, \texttt{basis.order} represents the number of functional principal components.

  \item[\texttt{bs\_degree}:]
  Specifies the degree of the piecewise polynomials for the B-spline basis; this argument is only necessary when using the B-spline basis.
\end{description}

\paragraph{Returned object:}  
The function returns an object of class \class{ME.fcQR\_CLS} (a list) containing the following elements:

\begin{enumerate}
    \item \textbf{\code{estimated\_beta\_hat}:} Estimated coefficients from the corrected loss function (including the functional component).
    \item \textbf{\code{estimated\_beta\_t}:} The estimated functional curve.
    \item \textbf{\code{SE\_est}:} The estimated parametric variance (returned only if \code{observed\_X} is not \code{NULL}).
    \item \textbf{\code{estimated\_Xbasis}:} The basis matrix used in the estimation.
    \item \textbf{\code{res\_naive}:} Results from the naive (uncorrected) method.
\end{enumerate}
   
% \textbf{Example codes}:
% \begin{CodeChunk}
% \begin{CodeInput}
%     rm(list = ls())
%     data(MECfda.data.sim.0.1)
%     res = ME.fcQR_CLS(data.Y = MECfda.data.sim.0.1$Y,
%                       data.W = MECfda.data.sim.0.1$W,
%                       data.Z = MECfda.data.sim.0.1$Z,
%                       tau = 0.5,
%                       grid_k = 4:7,
%                       grid_h = 1:2)
% \end{CodeInput}
% \end{CodeChunk}

\subsection*{Function \fct{ME.fcLR\_IV}}

\begin{CodeChunk}
\begin{CodeInput}
ME.fcLR_IV(
  data.Y,
  data.W,
  data.M,
  t_interval = c(0, 1),
  t_points = NULL,
  CI.bootstrap = F
)
\end{CodeInput}
\end{CodeChunk}

The \pkg{MECfda} package provides the function \fct{ME.fcLR\_IV} for implementing a bias-correction estimation method. The arguments are as follows:

\paragraph{Input data:}
\begin{itemize}
    \item \textbf{\code{data.Y} (Response Variable):}  
    The response variable can be provided as an atomic vector, a one-column matrix, or a data frame. The recommended format is a one-column data frame with a column name.

    \item \textbf{\code{data.W} (Measurement Data for $W$):}  
    This argument represents $W$, the observed measurement of $X$ in the statistical model. It should be provided as a data frame or matrix, where each row represents a subject and each column corresponds to a measurement (time) point.

    \item \textbf{\code{data.M} (Instrumental Variable):}  
    This argument provides the data for $M$, the instrumental variable. It should be provided as a data frame or matrix, with each row representing a subject and each column representing a measurement (time) point.
\end{itemize}

\paragraph{Key arguments:}
\begin{itemize}
    \item \textbf{\code{t\_interval} (Domain of the Functional Covariate):}  
    A two-element vector that specifies the domain (interval) of the functional covariate. The default is \code{c(0,1)}, representing the interval $[0,1]$.

    \item \textbf{\code{t\_points} (Measurement Time Points):}  
    Specifies the sequence of measurement (time) points. The default is \code{NULL}.

    \item \textbf{\code{CI.bootstrap} (Bootstrap Confidence Interval):}  
    A logical flag indicating whether to return a confidence interval using the bootstrap method. The default is \code{FALSE}.
\end{itemize}

\paragraph{Returned object:}  
The function returns an object of class \class{ME.fcLR\_IV}, which is a list containing the following elements:

\begin{enumerate}
    \item \textbf{\code{beta\_tW}:} Parameter estimates.
    \item \textbf{\code{CI}:} Confidence interval, which is returned only if \code{CI.bootstrap} is \code{TRUE}.
\end{enumerate}

% \textbf{Example codes}:
% \begin{CodeChunk}
% \begin{CodeInput}
%     rm(list = ls())
%     data(MECfda.data.sim.0.3)
%     res = ME.fcLR_IV(data.Y = MECfda.data.sim.0.3$Y,
%                      data.W = MECfda.data.sim.0.3$W,
%                      data.M = MECfda.data.sim.0.3$M)
% \end{CodeInput}
% \end{CodeChunk}

\section{R code for example project}\label{Example project codes}

\begin{CodeChunk}
\begin{CodeInput}
library(nhanesA)
library(dplyr)
library(tidyr)
library(readr)

rm(list = ls())

#### Download and preprocess data ####

## Download hourly-level Physical Activity Monitor data for 2011–2012
PAXHR = nhanes("PAXHR_G")

## Add a variable indicating the hour of the day (0–23)
PAXHR = PAXHR %>%
  dplyr::group_by(SEQN, PAXDAYH) %>%
  dplyr::arrange(PAXSSNHP, .by_group = TRUE) %>%
  dplyr::mutate(hour_in_day = row_number() - 1) %>%
  dplyr::ungroup()

## Identify subjects with data available for all 9 days (1–9)
id_all9days = PAXHR %>%
  dplyr::group_by(SEQN) %>%
  dplyr::summarise(n_days = n_distinct(PAXDAYH), .groups = "keep") %>%
  dplyr::filter(n_days == 9) %>%
  dplyr::pull(SEQN)

## Filter data to include only those subjects
PAXHR = PAXHR %>%
  dplyr::filter(SEQN %in% id_all9days)
rm(id_all9days)



## Screen out the individuals who have 24-hour complete data every day 
##   from the second to the eighth day

## Identify subjects who have exactly 24 hourly records for each of days 2 to 8
id_completeData = PAXHR %>%
  dplyr::filter(as.integer(PAXDAYH) %in% 2:8) %>%  # Days 2–8 only
  dplyr::group_by(SEQN, PAXDAYH) %>%              # Group by subject and day
  dplyr::summarise(n_hour = n_distinct(PAXSSNHP), .groups = "drop") %>%
  dplyr::filter(n_hour == 24) %>%                 # Keep only full 24-hour days
  dplyr::group_by(SEQN) %>%
  dplyr::summarise(n_day = n()) %>%               # Count number of such days
  dplyr::filter(n_day == 7) %>%                   # Keep only if all 7 days (2–8) are complete
  dplyr::pull(SEQN)

## Keep only selected subjects and days 2–8
PAXHR = PAXHR %>%
  filter(SEQN %in% id_completeData, as.integer(PAXDAYH) %in% 2:8)
PAXHR$PAXDAYH = droplevels(PAXHR$PAXDAYH)

## Keep only relevant variables
PAXHR = PAXHR[, colnames(PAXHR) %in% c('SEQN','PAXDAYH','hour_in_day','PAXMTSH')]

## Reshape to wide format: 24 hourly columns
PAXHR = PAXHR %>%
  tidyr::pivot_wider(
    names_from = hour_in_day,
    values_from = PAXMTSH
  )

## Reorder rows and split by day into 3D array: subject × hour × day
PAXHR = PAXHR[order(match(PAXHR$SEQN, id_completeData)), ]
day_list_PA = split(PAXHR[, as.character(0:23)], PAXHR$PAXDAYH)
day_list_PA = lapply(day_list_PA, as.matrix)
PA_array = abind::abind(day_list_PA, along = 3)

## Download demographic data
DEMO = nhanes("DEMO_G")
DEMO = DEMO[DEMO$SEQN %in% id_completeData, c('SEQN','RIAGENDR','RIDAGEYR','RIDRETH1')]
DEMO = DEMO[order(match(DEMO$SEQN, id_completeData)), ]
DEMO = DEMO[, c('RIAGENDR','RIDAGEYR','RIDRETH1')]

## Download the body measurement data and keep the variable BMI
BMX = nhanes("BMX_G")
BMX = BMX[BMX$SEQN %in% id_completeData, c('SEQN','BMXBMI')]
BMX = BMX[order(match(BMX$SEQN, id_completeData)), ]
BMX = BMX[, c('BMXBMI'), drop = FALSE]



## Download step count data (minute-level, 1440 minutes per day)
destfile = "nhanes_1440_adeptsteps.csv.xz"
url = paste0("https://physionet.org/files/minute-level-step-count-nhanes/",
              "1.0.1/csv/nhanes_1440_adeptsteps.csv.xz")
download.file(url, destfile,
              method = "curl", extra = "--retry 10 --max-time 600")

## Read step count data and subset to selected subjects and days
steps_data = readr::read_csv(destfile)
steps_data = steps_data[steps_data$SEQN %in% id_completeData, ]
steps_data = steps_data[steps_data$PAXDAYM %in% 2:8, ]
steps_data = steps_data[order(match(steps_data$SEQN, id_completeData)), ]

## Collapse 1440-minute data into 24 hourly sums
stepCounts = as.matrix(steps_data[, -(1:3)])
col_groups = ceiling(seq_along(1:1440) / 60)
stepCounts = sapply(unique(col_groups), function(i) rowSums(stepCounts[, col_groups == i]))
stepCounts = as.data.frame(stepCounts)
colnames(stepCounts) = paste('hr', 0:23, sep = '_')
stepCounts = cbind(steps_data[, c("SEQN", "PAXDAYM")], stepCounts)

## Convert to 3D array: subject × hour × day
day_list_SC = split(stepCounts[, paste('hr', 0:23, sep = '_')], stepCounts$PAXDAYM)
day_list_SC = lapply(day_list_SC, as.matrix)
SC_array = abind::abind(day_list_SC, along = 3)

## Remove subjects with missing values in any array or covariates
have.missing =
  apply(PA_array, 1, anyNA) |
  apply(SC_array, 1, anyNA) |
  apply(BMX, 1, anyNA) |
  apply(DEMO, 1, anyNA)

PA_array = PA_array[!have.missing,,]
SC_array = SC_array[!have.missing,,]
BMX = BMX[!have.missing,, drop = FALSE]
DEMO = DEMO[!have.missing,, drop = FALSE]
id_completeData = id_completeData[!have.missing]

## Clean up environment
rm(list = setdiff(ls(), c('PA_array','SC_array','DEMO','BMX')))



#### Visualization: plot average activity on day 2 ####
library(ggplot2)


pdf("average_activity_second_day.pdf", width = 7, height = 4)

ggplot(data.frame(hour = 0:23,
                  activity = colMeans(PA_array[,,1])), 
       aes(x = hour, y = activity)) +
  geom_line(size = 1) +
  geom_point(size = 2) +
  scale_x_continuous(breaks = hours) +
  labs(
    title    = "Average Activity Intensity Across 24 Hours (Day 2)",
    subtitle = "Unit: MIMS-unit",
    x        = "Hour of Day",
    y        = "Average Activity Intensity (MIMS-unit)"
  ) +
  theme_minimal(base_size = 12) +
  theme(
    plot.title   = element_text(hjust = 0.5, face = "bold"),
    plot.subtitle= element_text(hjust = 0.5),
    panel.grid.major.x = element_blank(),
    panel.grid.minor = element_blank()
  )

dev.off()



#### Analysis using MECfda package ####

## Use a small subset of 100 subjects for demonstration
set.seed(4)
ind_sub = sample(1:nrow(DEMO), 100, replace = FALSE)
SC_array  = SC_array[ind_sub,,]
PA_array  = PA_array[ind_sub,,]
BMX = BMX[ind_sub,, drop = FALSE]
DEMO = DEMO[ind_sub,, drop = FALSE]



library(MECfda)

## Functional linear regression
PA_day2 = PA_array[,,1]
res1 = fcRegression(
  Y = BMX,
  FC = functional_variable(PA_day2, 0, 24, 0:23),
  Z = DEMO,
  basis.order = 5, basis.type = c('Bspline'))

## Functional quantile regression
res2 = fcQR(
  Y = BMX,
  FC = functional_variable(PA_day2, 0, 24, 0:23),
  Z = DEMO,
  basis.order = 5, basis.type = c('Bspline'))


## Measurement error corrected functional regression

## MEM method
res3 = ME.fcRegression_MEM(
  data.Y = BMX,
  data.W = PA_array,
  data.Z = DEMO,
  t_interval = c(0, 24),
  t_points = 0:23,
  method = 'UP_MEM',
  family.W = "gaussian",
  basis.order = 5L,
  bs_degree = 3L,
  basis.type = 'Bspline')


SC_day2 = SC_array[,,1]
## Instrumental variable + SIMEX quantile SoFR regression
res4 = ME.fcQR_IV.SIMEX(
  data.Y = BMX,
  data.W = PA_day2,
  data.Z = DEMO,
  data.M = SC_day2,
  t_interval = c(0, 24),
  t_points = 0:23,
  basis.order = 5L,
  bs_degree = 3L,
  tau = 0.5,
  basis.type = 'Bspline')


## Corrected Loss Score Method for quantile SoFR
## Convert non-numeric covariates to design matrix
## right now ME.fcQR_CLS only allow numeric matrix as argument data.Z
DEMO_des = model.matrix(~ . , data = DEMO)
DEMO_des = DEMO_des[,-1,drop = FALSE]
res5 = ME.fcQR_CLS(
  data.Y = BMX,
  data.W = PA_array,
  data.Z = DEMO_des,
  t_interval = c(0, 24),
  t_points = 0:23,
  tau = 0.5,
  grid_k = 4:7,
  grid_h = 1:2)


## Instrumental variable linear regression
res6 = ME.fcLR_IV(
  data.Y = BMX,
  data.W = PA_day2,
  data.M = SC_day2)

## Plot results
plot(res1$FC.BasisCoefficient[[1]])
plot(res2$FC.BasisCoefficient[[1]])
plot(res3$FC.BasisCoefficient[[1]])
plot(res4$coef.X)
plot(0:23, res5$estimated_beta_t, type = 'l')
plot(0:23,res6$beta_tW, type = 'l')

## Define interpolation + extrapolation function
interp_extrap_fun <- function(x, y) {
  stopifnot(length(x) == length(y))
  ord <- order(x)
  x <- x[ord]; y <- y[ord]
  
  f_interp <- splinefun(x, y, method = "monoH.FC")
  
  left_slope  <- (y[2] - y[1]) / (x[2] - x[1])
  right_slope <- (y[length(y)] - y[length(y) - 1]) / (x[length(x)] - x[length(x) - 1])
  
  function(x0) {
    res <- numeric(length(x0))
    res[x0 < x[1]] <- y[1] + left_slope * (x0[x0 < x[1]] - x[1])
    res[x0 > x[length(x)]] <- y[length(y)] + 
      right_slope * (x0[x0 > x[length(x)]] - x[length(x)])
    res[x0 >= x[1] & x0 <= x[length(x)]] <- f_interp(x0[x0 >= x[1] & x0 <= x[length(x)]])
    
    return(res)
  }
}


## Define a ggplot2-based plotting function for \hat{\beta}(t)
plot_beta_curve <- function(t, beta, method = NULL,
                            line_color = "black", line_size = 1, width = 6, height = 4) {
  stopifnot(length(t) == length(beta))
  
  filename = paste0(method,".pdf")
  
  library(ggplot2)
  
  data <- data.frame(
    time = t,
    beta = beta
  )
  
  p <- ggplot(data, aes(x = time, y = beta)) +
    geom_line(color = line_color, size = line_size) +
    labs(
      x = "Time of day (hour)",
      y = expression(hat(beta)(t)),
      title = if (!is.null(method)) paste("Estimated", expression(beta(t)), "using", method) else NULL
    ) +
    theme_minimal(base_size = 12) +
    theme(
      plot.title = element_text(hjust = 0.5, face = "bold"),
      axis.title = element_text(face = "bold"),
      panel.grid.minor = element_blank()
    ) +
    scale_x_continuous(breaks = seq(0, 24, by = 4), limits = c(0, 24))
  
  ggsave(filename, plot = p, width = width, height = height, units = "in")
  message("Plot saved to ", filename)
}


## Plot estimated curves
t = seq(0,24,0.1)
plot_beta_curve(t,basis2fun(res1$FC.BasisCoefficient[[1]],t), "fcRegression")
plot_beta_curve(t,basis2fun(res2$FC.BasisCoefficient[[1]],t), "fcQR")
plot_beta_curve(t,basis2fun(res3$FC.BasisCoefficient[[1]],t), "ME.fcRegression_MEM")
plot_beta_curve(t,basis2fun(res4$coef.X,t), "ME.fcQR_IV.SIMEX")
plot_beta_curve(t,interp_extrap_fun(0:23, res5$estimated_beta_t)(t), "ME.fcQR_CLS")
plot_beta_curve(t,interp_extrap_fun(0:23, res6$beta_tW)(t), "ME.fcLR_IV")

q(save = 'no')
\end{CodeInput}
\end{CodeChunk}

\end{appendix}

\end{document}